\documentclass[twocolumn]{aastex61}
\pdfoutput=1 
\usepackage{amsmath,amstext,amssymb}
\usepackage[T1]{fontenc}
\usepackage{apjfonts} 
\usepackage[figure,figure*]{hypcap}
\usepackage{todonotes}
\usepackage{verbatim}
\usepackage{ulem}

\newcommand{\St}{\mathrm{St}} 

\shorttitle{Tracking dust grains}
\shortauthors{Misener et al.}

\begin{document}

\title{Tracking Dust Grains During Transport and Growth in Protoplanetary Disks}

\author{William Misener}
\affiliation{Department of Earth, Planetary, and Space Sciences, The University of California, Los Angeles, 595 Charles E. Young Drive East, Los Angeles, CA 90095, USA}
\affiliation{Department of the Geophysical Sciences, The University of Chicago, 5734 South Ellis Avenue, Chicago, IL 60637, USA}

\author{Sebastiaan Krijt}
\affiliation{Department of Astronomy/Steward Observatory, The University of Arizona, 933 North Cherry Avenue, Tucson, AZ 85721, USA}
\affiliation{Earths in Other Solar Systems Team, NASA Nexus for Exoplanet System Science}
\affiliation{Hubble Fellow}
\author{Fred J Ciesla}
\affiliation{Department of the Geophysical Sciences, The University of Chicago, 5734 South Ellis Avenue, Chicago, IL 60637, USA}
\affiliation{Earths in Other Solar Systems Team, NASA Nexus for Exoplanet System Science}


\begin{abstract}
Protoplanetary disks are dynamic objects, within which dust grains and gas are expected to be redistributed over large distances.  Evidence for this redistribution is seen both in other protoplanetary disks and in our own Solar System, with high-temperature materials thought to originate close to the central star found in the cold, outer regions of the disks.  While models have shown this redistribution is possible through a variety of mechanisms, these models have generally ignored the possible growth of solids via grain-grain collisions that would occur during transit.  Here we investigate the interplay of coagulation and radial and vertical transport of solids in protoplanetary disks, considering cases where growth is limited by bouncing or by fragmentation.  We find that in all cases, growth effectively limits the ability for materials to be carried outward or preserved at large distances from the star.  This is due to solids being incorporated into large aggregates which drift inwards rapidly under the effects of gas drag. We discuss the implications for mixing in protoplanetary disks, and how the preservation of high temperature materials in outer disks may require structures or outward flow patterns to avoid them being lost via radial drift.
\end{abstract}


\section{Introduction}
The earliest stages of planet formation occur within protoplanetary disks, the 
predominantly gaseous clouds that surround young stars immediately after their birth. 
Within these disks, fine dust begins to coagulate, growing from individual grains that are suspended within the gas, to pebbles which begin to settle to the disk midplane and drift inwards, to planetesimals which follow their own orbits around the star without much regard to their interactions with the gas.  The details of this growth remain 
enigmatic, though it is generally accepted that the early stages of growth involve low
velocity collisions that allow dust grains to stick to one another \citep[e.g.,][]{dominiktielens1997}.  As
grains grow larger, the collisions become more energetic, with bouncing, erosion, and
fragmentation becoming more likely \citep{blumwurm2008}.  This typically occurs once aggregates enter the
pebble regime, where they are large enough (millimeters to meters) that they begin to decouple from the gas.  At this stage, growth by sticking is likely to be frustrated \citep{guttler2010,zsom2010},  leaving the formation of larger bodies to processes such as turbulent 
concentration \citep[e.g.][]{cuzzi01,cuzzi08} or the streaming instability 
\citep[e.g.][]{youdin05,johansen07,carrera15}.

If and when collisional growth is frustrated will depend on many factors, including 
the sizes,  structures, and  compositions of the dust grains and aggregates.  Most
important in the outcome, however, will be the velocities at which the dust grains 
strike one another.  Grains that are meter-sized or smaller will have their
motions,  and thus their relative velocities, controlled by their interactions with the 
gas.  The key dynamic processes expected to be at play are vertical settling to the midplane,
inward radial migration due to gas drag, and turbulent diffusion.  The relative importance of
each will depend on the location in the disk and the sizes of the grains of interest: small grains in dense regions are well-coupled
to the gas and thus are most strongly affected by turbulent diffusion, while larger
particles have their motions largely controlled by vertical settling and
radial drift \citep[e.g.,][]{dominik2007,birnstiel2010}.

While these effects have been considered in previous coagulation models to establish relative velocities, collision rates, and collision outcomes \citep[e.g.][]{weidenschilling97,ciesla07b,birnstiel09,pinilla12,krijtcieslabergin2016}, an important consequence of these dynamics is that grains will constantly be in motion, moving from one region of the
disk to another as they grow. Turbulent diffusion, in particular, can carry grains throughout
the disk, lofting them to high altitudes or allowing them to migrate outward away from the star,
offsetting the effects of vertical settling  or radial drift.  In fact, outward diffusion has been invoked, at least
in part, to explain the presence of high-temperature grains in comets 
\citep{bockelee02,ciesla07}, the preservation of CAIs in chondritic meteorites 
\citep{cuzzi03,ciesla2010,jacquet11,desch2018}, and the redistribution of water with different
D/H ratios \citep{mousis00,yang13}.  These studies showed that particles of a given size could be preserved and redistributed  throughout the lifetime of the disk despite other effects which would cause them to be lost to the Sun.   

While such transport of individual grains is expected within a protoplanetary disk, growth and fragmentation of solids will also readily occur as  grains and aggregates encounter one another, an effect that is often ignored in these transport models.  In the denser regions of the disk, the timescales for these encounters can be very short \citep[$\sim$10$^{2}$-10$^{4}$ yrs, see][]{birnstiel09,pinilla12,krijtciesla2016} 
relative to the lifetimes of disks and the timescales over which materials may be transported ($\sim$10$^{6}$ yrs).  As a result, these particles should not remain in aggregates or solids of a fixed size throughout their time in the disk; instead their sizes, and the dynamical processes which control their movement through the disk, will evolve throughout.

The interplay of transport and dust growth for individual particles was investigated by 
\cite{krijtciesla2016} when considering the vertical motions of
materials and the exchange between the midplane and surface of a disk.  While small 
particles would normally readily diffuse between these two regions, it was found that 
interactions with other dust (i.e. growth) would hinder this process, limiting the ability of midplane materials to reach 
the disk surface.  As small grains encountered other solids as they were lofted to higher altitudes, they would tend to stick, 
get incorporated into larger bodies, and thus settle back toward the midplane again.  
The importance of this effect varied with the dust-to-gas ratio, but for canonical 
conditions, individual grains were largely trapped around the midplane, 
as they were more likely to be incorporated into larger bodies that settled downward than diffuse up to the disk surface. 

Here we investigate how dust dynamics and growth combine to affect the radial redistribution of materials  within 
 a protoplanetary disk. In the next section we describe the basic framework within which we consider dust evolution in the protoplanetary disk.  In Section \ref{sec:smallgrowth}, a simple picture for dust growth that highlights the effects that will control how grains evolve individually and as part of larger aggregates is introduced to illustrate key behaviors and outcomes.  In Section \ref{sec:bouncing}, we develop a more realistic model for growth of dust grains to sizes that are set by the bouncing barrier.  In Section \ref{sec:fragmentation} we apply that same model to investigate the history of dust grains when fragmentation sets the limit on the sizes of dust growth.  In Section \ref{sec:parameterexp}, we discuss the effects that different model parameters have on our results, and we summarize our key findings and discuss the implications for protoplanetary disk dust evolution in Section \ref{sec:disc}.

\section{Physical processes}

\subsection{Disk model}\label{sec:diskmodel}

We consider an azimuthally symmetric protoplanetary disk around a solar mass star whose surface density and temperature profiles are described by power laws:
\begin{equation}
\Sigma \left(r \right) = \Sigma_0 \left(\frac{r}{R_{0}} \right)^{-p},
\end{equation}
\begin{equation}
T \left( r \right)= T_0  \left(\frac{r}{R_{0}} \right)^{-q},
\end{equation}
where $r$ is the radial location in the disk, $R_{0}$ is the  location where the reference surface density, $\Sigma_{0}$, and temperature, $T_{0}$, are defined, and $p$ and $q$ are the indices describing the change in the respective values. Here we assume $\Sigma_{0}=2000$ g cm$^{-2}$ and $T_{0}=150$ K at $R_{0}=1$ AU with $p=1$ and $q=3/7$ as expected for an irradiated disk \citep[e.g.][]{chianggoldreich97}, with the disk taken to be isothermal with height $z$ above the midplane. The impact of different gas surface density profiles on the results is discussed in Section~\ref{sec:parameterexp}.

We assume the disk is passive, such that there are no large-scale flows associated with mass or angular momentum transport.  As such, the properties of the disk remain constant throughout the simulation, an issue that we return to later in our discussion.  While the disk properties are unchanging, we do assume that there is some level of diffusion within the disk, which is characterized by the turbulent parameter $\alpha$ such that the gas has characteristic diffusivity, $D= \alpha c_{s} H$, where $H=c_{s}/\Omega$ is the local scale height,  $c_{s}$ is the local speed of sound, $\Omega$ the local Keplerian frequency, and $\alpha$ describes the level of turbulent mixing in the disk.  Here we adopt a value of $\alpha=10^{-4}$ for our canonical cases, comparable to the value used by \citet{desch2018} for the inner solar system and to the lower limits inferred for the outer regions of protoplanetary disks \citep{dullemond18}, though given the uncertainties on this parameter, we discuss the outcomes expected for various changes in this and other model parameters in Section \ref{sec:parameterexp}.

Given the surface density and temperature profiles above, the local gas density in the disk can be found as: 
\begin{equation}
\rho_\mathrm{g} \left( r, z \right) = \frac{\Sigma(r)}{\sqrt{2 \pi}H} \exp{\left( \frac{- z^{2}}{2 H^{2}} \right)}.
\end{equation}
This sets the local conditions around the dust grain and the evolution it experiences within those respective environments.

\subsection{Dust grain dynamics}\label{sec:dynamics}

Our goal is to understand the interplay of dust dynamics and growth within a protoplanetary disk, and as such, we must track the specific environments that grains move through as this will determine their ability to grow.  We therefore adopt the particle tracking methods of \citet{ciesla2010b} and \citet{ciesla2011}, which follow the movement of individual grains as they move within a disk, accounting for vertical settling, radial drift due to gas drag, and the diffusive motions of the particle (in both the radial and vertical directions) using a Monte Carlo approach. These methods have been shown to reproduce the expected collective dynamical behavior of trace gas and dust species within a diffusive protoplanetary disk, including the Gaussian distribution of materials about the disk midplane \citep{ciesla2010b}. Similar approaches were also used in \citet{zsom2011, ros2013, krijtciesla2016,krijtcieslabergin2016} to simulate the dynamics of particles of a variety of sizes.

The amount a spherical particle of radius $s$ and internal density $\rho_{\bullet}$ moves in a given timestep is set by the local conditions in the disk, as well as the Stokes number of the particle,  $\St=t_\mathrm{s} \Omega$, where 
\begin{equation}
t_{\mathrm{s}} = \begin{cases} \dfrac{\rho_{\bullet} s}{\rho_{\mathrm{g}} v_{\mathrm{th}}} &\mbox{if } s \leq \dfrac{9}{4} \lambda_{\mathrm{mfp}} \vspace{3mm}\\
\dfrac{4s}{9 \lambda_{\mathrm{mfp}}} \dfrac{\rho_{\bullet} s}{\rho_{\mathrm{g}} v_{\mathrm{th}}} &\mbox{if } s > \dfrac{9}{4} \lambda_{\mathrm{mfp}} \end{cases} 
\end{equation}
is the particle stopping time at the particular location in the disk. Here, $v_{\mathrm{th}}= (8 k_{\mathrm{B}} T/ (\pi \mu m_{\mathrm{p}}))^{1/2}$ and $\lambda_{\mathrm{mfp}}=\mu m_{\mathrm{p}}/(\sigma_{\mathrm{mol}} \rho_{\mathrm{g}})$ are the local gas thermal speed and mean free path respectively \citep{okuzumi2012}. This gives a diffusivity of the particles $D_\mathrm{p} = D / ( 1+ \St^{2} )$ \citep{youdin2007}, which is used in determining the 
 random displacements associated with their motions in the turbulent environment.  

\begin{figure*}[t]
\centering
\includegraphics[width=.5\textwidth]{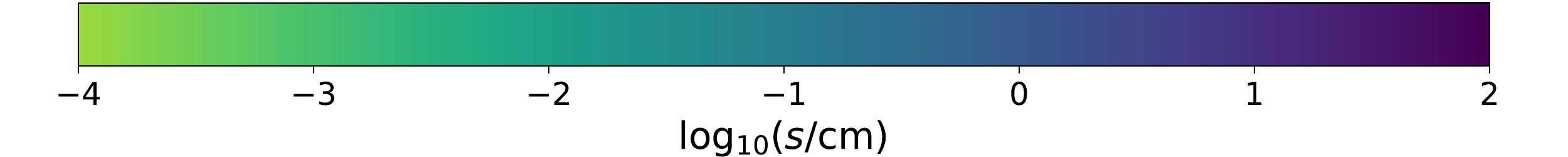}
\\
\includegraphics[width=.32\textwidth]{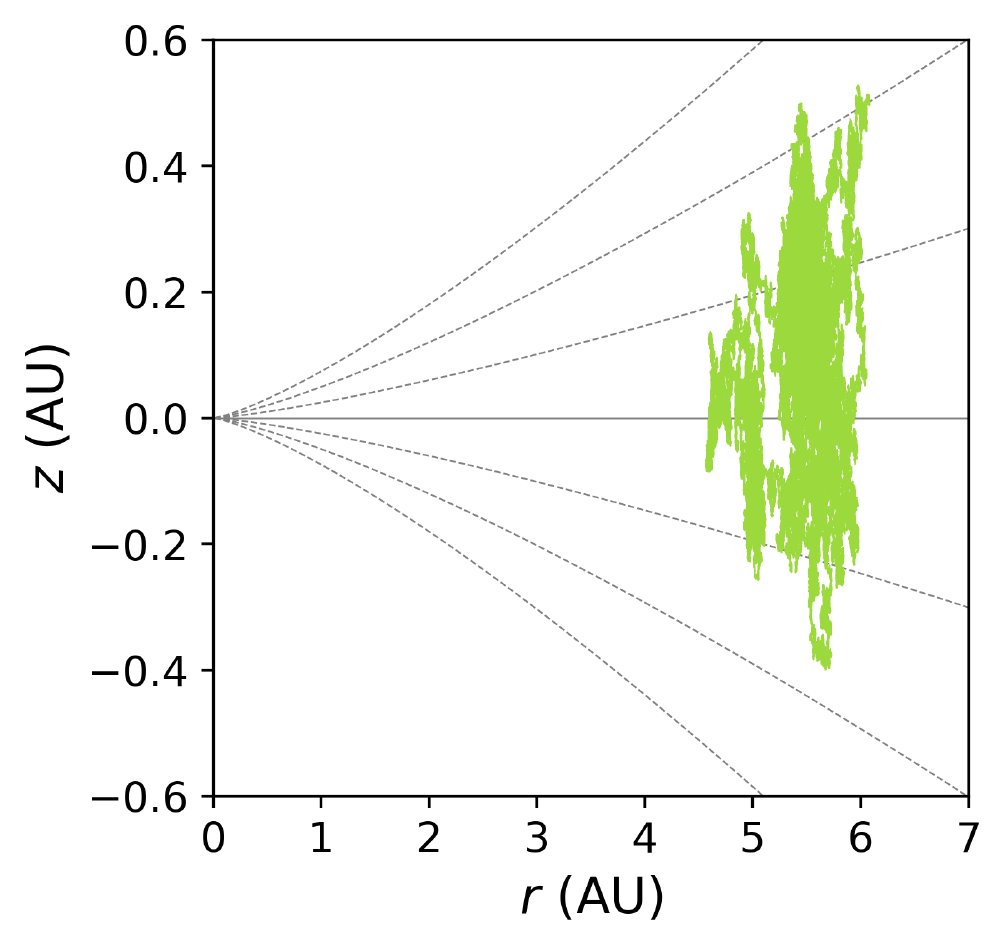}
\includegraphics[width=.32\textwidth]{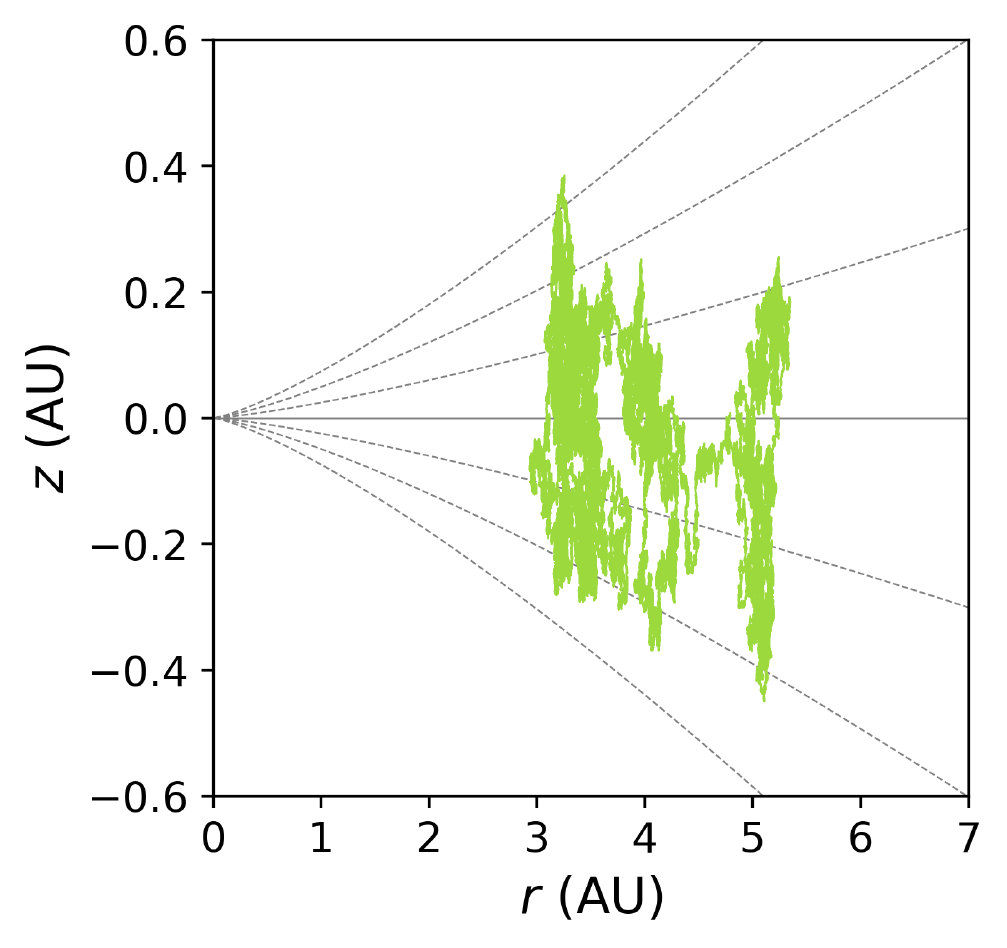}
\includegraphics[width=.32\textwidth]{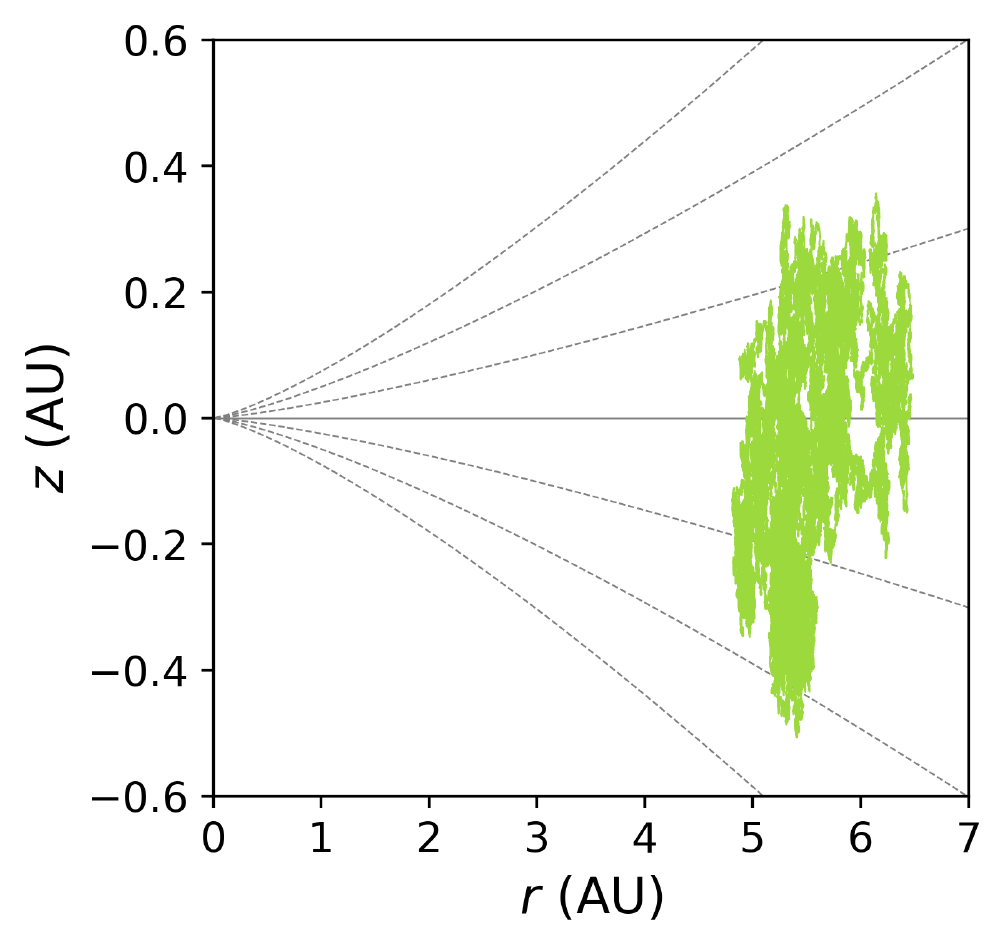}
\includegraphics[width=.32\textwidth]{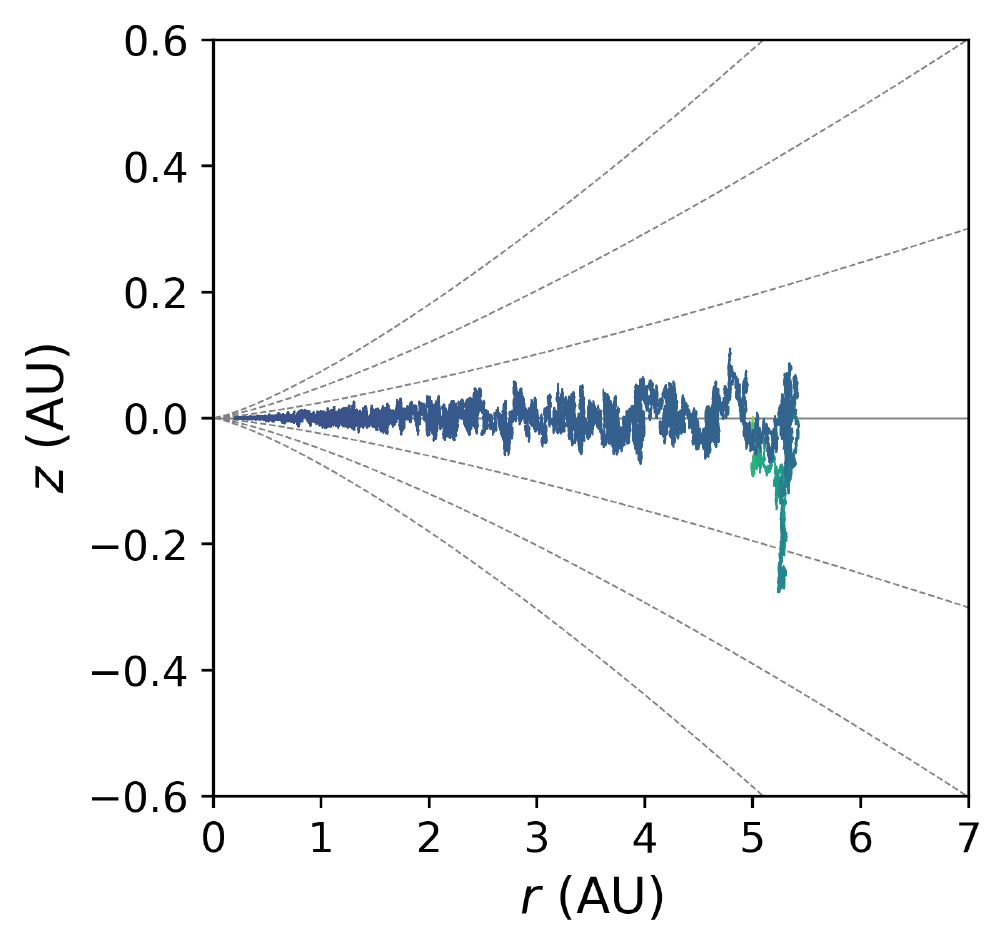}
\includegraphics[width=.32\textwidth]{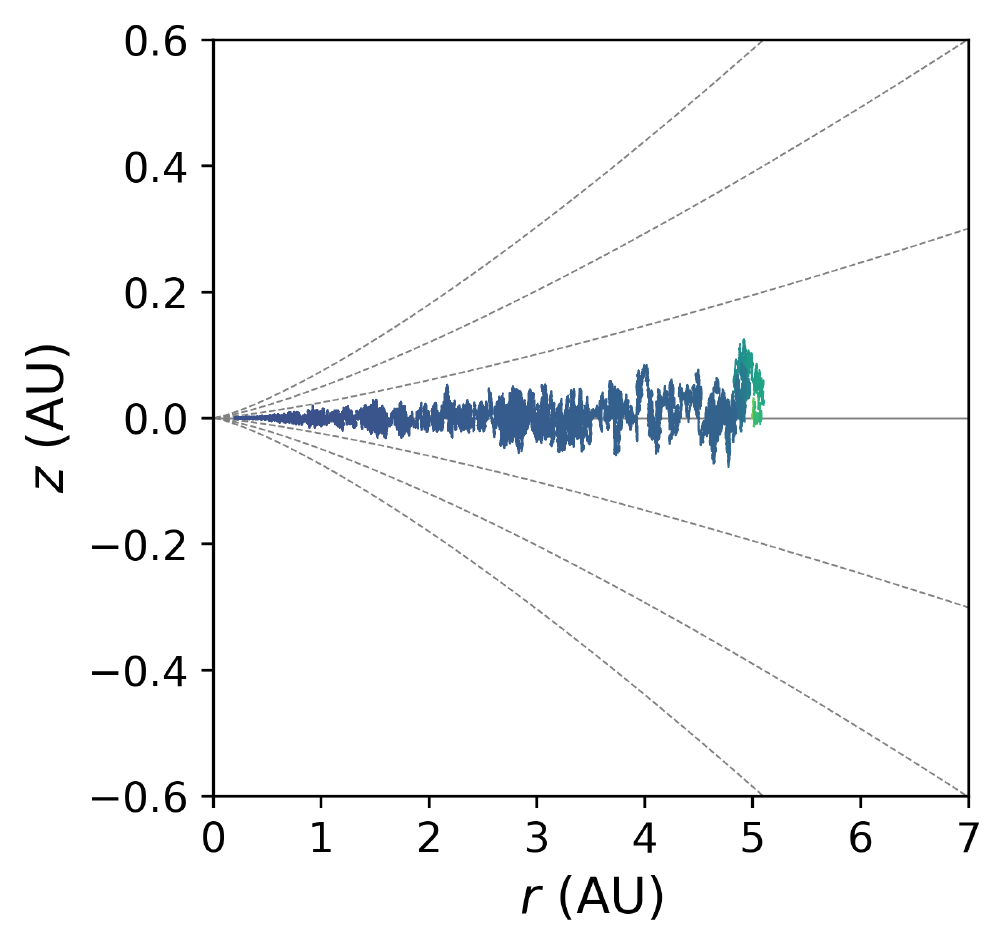}
\includegraphics[width=.32\textwidth]{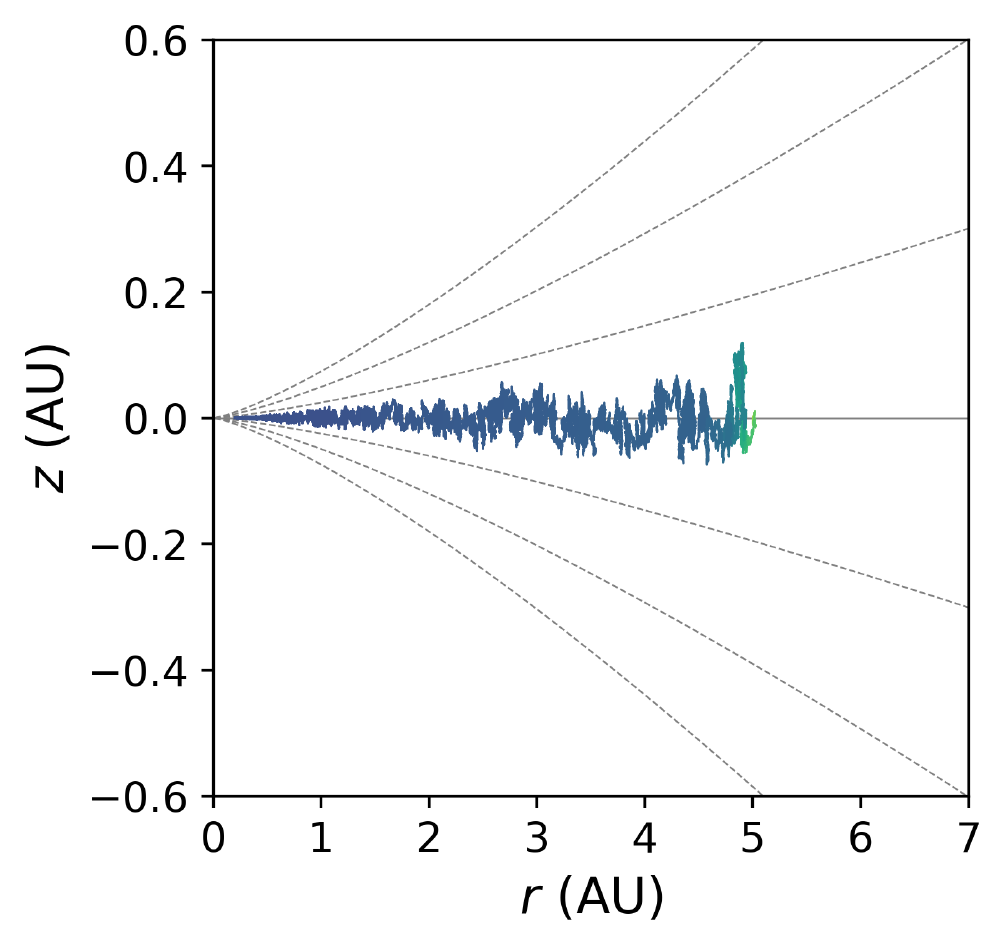}\\
\includegraphics[width=.32\textwidth]{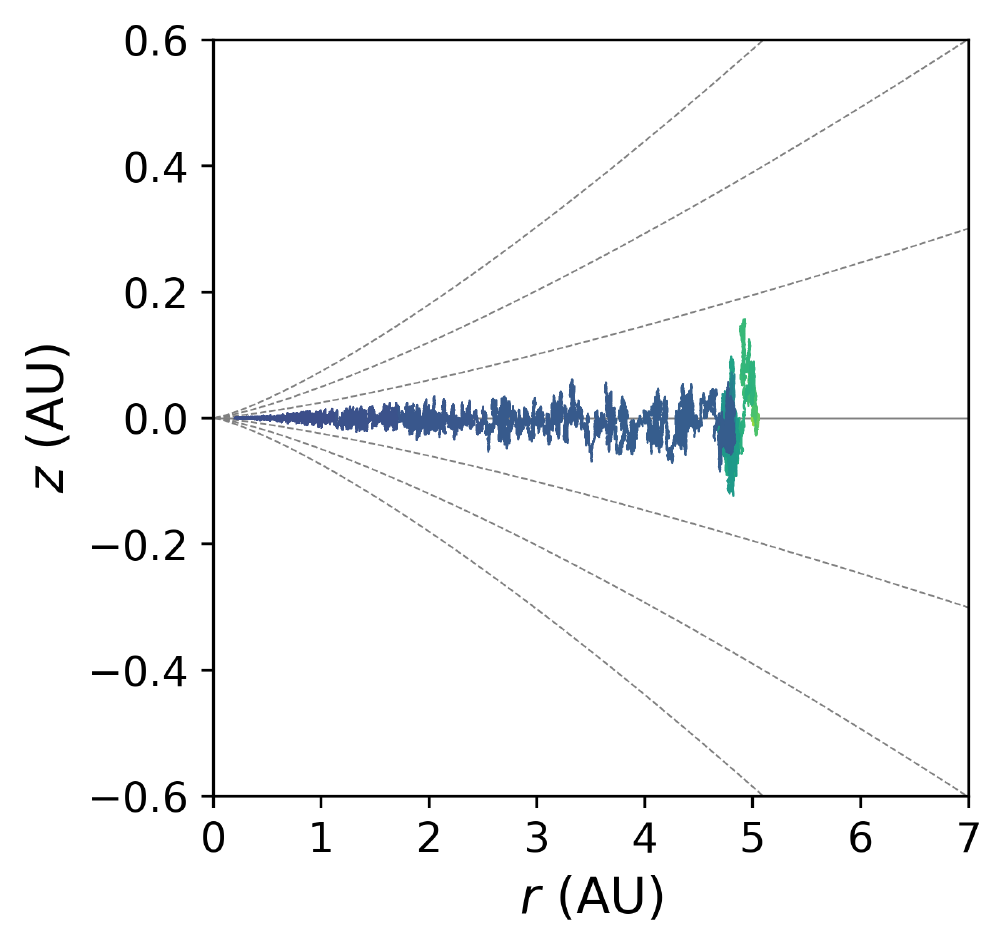}
\includegraphics[width=.32\textwidth]{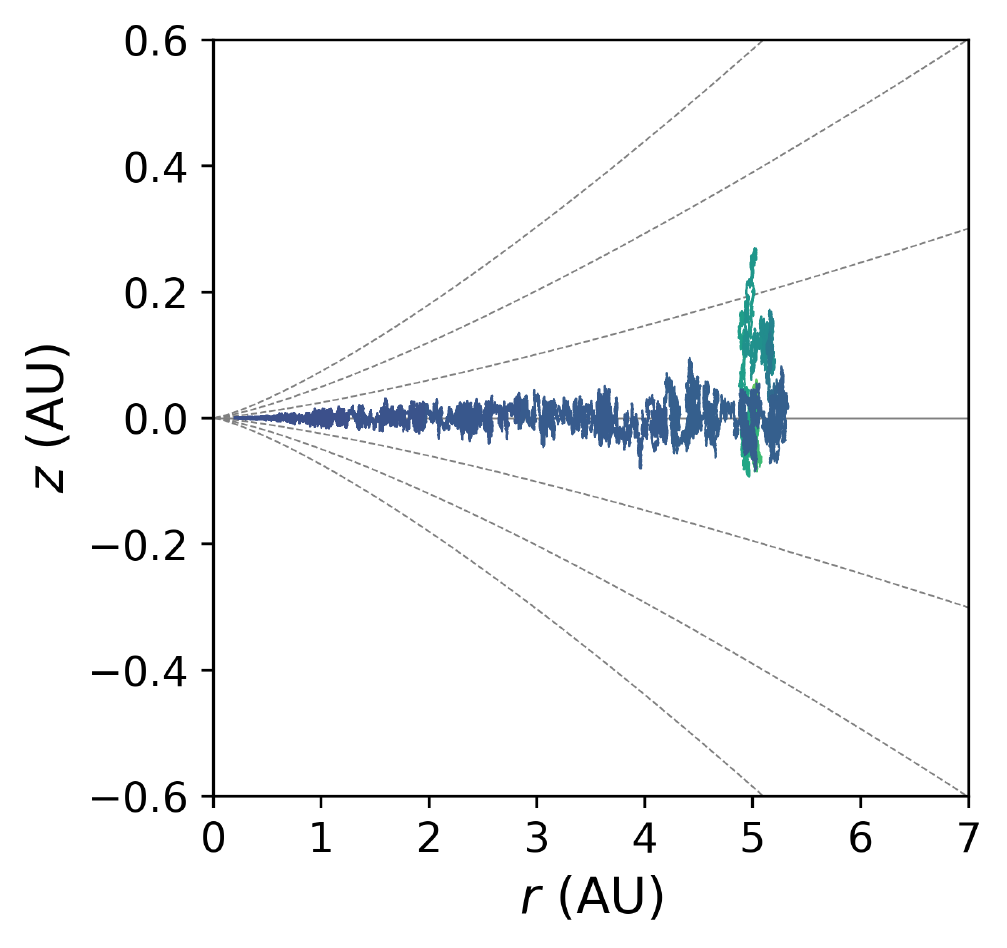}
\includegraphics[width=.32\textwidth]{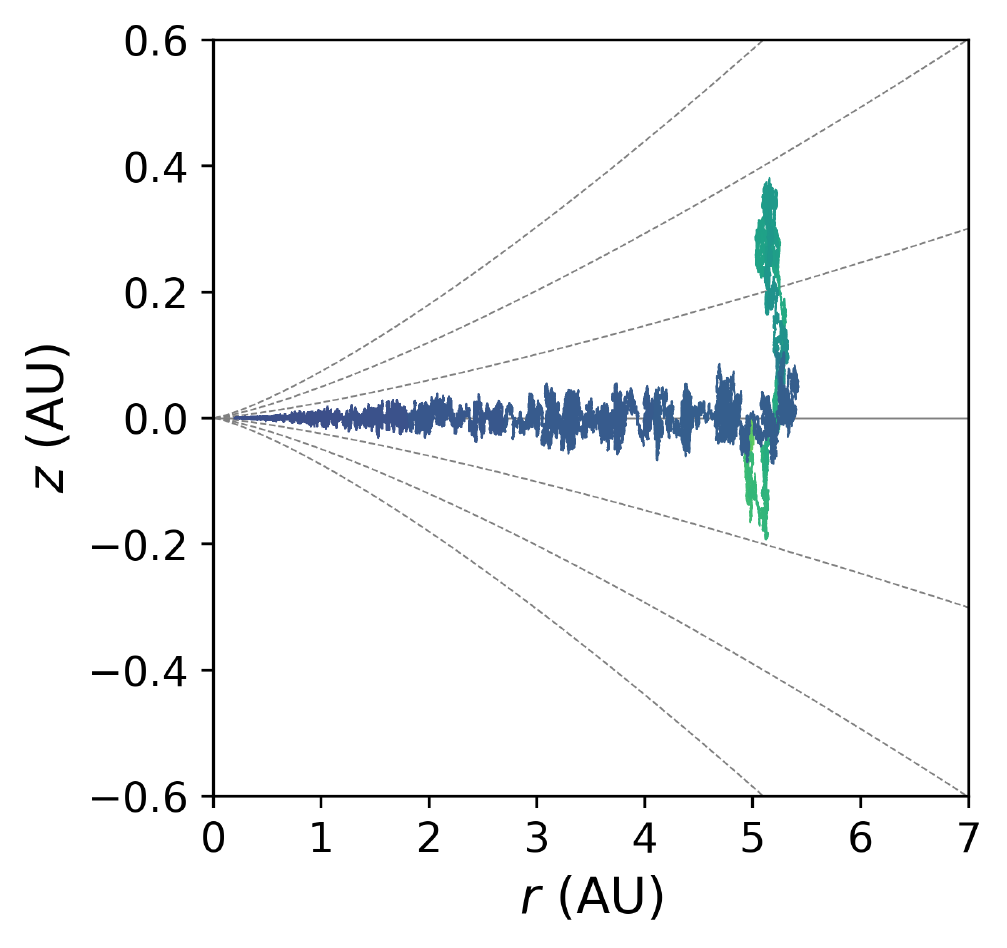}

\caption{Paths through the disk of individual micron-size dust grains released at $r=5$ AU, $z=0$. Colors indicate the size of the grain or of the aggregate it gets incorporated into. Dashed lines show $z/H=\pm1,2,3$. Top row: No growth scenario; the particle size does not increase. Middle row: Sweep-up scenario (Sect.~\ref{sec:smallgrowth}). Bottom row: Mixed growth to the bouncing barrier (Sect.~\ref{sec:bouncing}).}\label{fig:individual}
\end{figure*}

The top row of Fig.~\ref{fig:individual} shows examples of the trajectories that individual $s=1~\mathrm{\mu m}$ grains would take through the disk over a period of $3 \times 10^{5}$ years.  Here we have assumed that the particles do not interact with other grains over the course of the simulation, remaining the same size throughout.  The particles are assumed to have an internal density of $\rho_{\bullet}=$1.5 g cm$^{-3}$ and are released at the disk midplane at a distance $r=5\mathrm{~AU}$ from the star, where they have a Stokes number of $\St \sim 10^{-6}$. At such low Stokes numbers, the particles are largely coupled to the gas, and thus their motions are dominated by stochastic turbulent diffusion, with settling and drift having minimal effects.  As illustrated by the various trajectories shown in Fig.~\ref{fig:individual}, every particle follows its own, unique path through the disk, being exposed to its own series of environments.  Such pathways are important to understand, as the integrated effects of each environment will determine the final chemical properties of the solid \citep{cieslasandford12}.

\subsection{Dust coagulation}
Rather than remaining small and individual objects, dust grains are expected to frequently collide with other dust particles, becoming part of increasingly larger aggregates (and aggregates of aggregates) as long as collisions are gentle enough to result in sticking \citep[e.g.,][]{brauer2008,kothe2013}. In more energetic impacts, and depending on the mass ratio and structure of the colliders, a variety of outcomes including bouncing, erosion, and catastrophic fragmentation become more likely \citep[e.g.,][]{blumwurm2008,guttler2010}. In such events, grains that are initially part of a large aggregate can either remain there (in the case of bouncing) or be ejected as a smaller piece (in the case of fragmentation). With the size of the aggregate containing any particular grain changing frequently and sometimes dramatically, the dynamical behavior of grains also changes in time, reflecting the aerodynamical properties of their host aggregate. 

While the interplay between coagulation and radial transport through a disk has been considered in vertically-integrated models \citep{birnstiel09, pinilla12, okuzumi2012}, the grid-based approach of such studies is not well suited for studying detailed trajectories of individual grains. Alternatively, representative particle models for coagulation have also been developed\footnote{See \citet{drazkowska2014} for a comparison of both methods.} that allow tracking of dust particles and their properties (e.g., porosity, ice content) \citep{ormel2007,zsom08,krijtcieslabergin2016}; however, these have been applied only to localized regions of the disk.

\begin{figure*}[t]
\centering
\includegraphics[width=.85\textwidth]{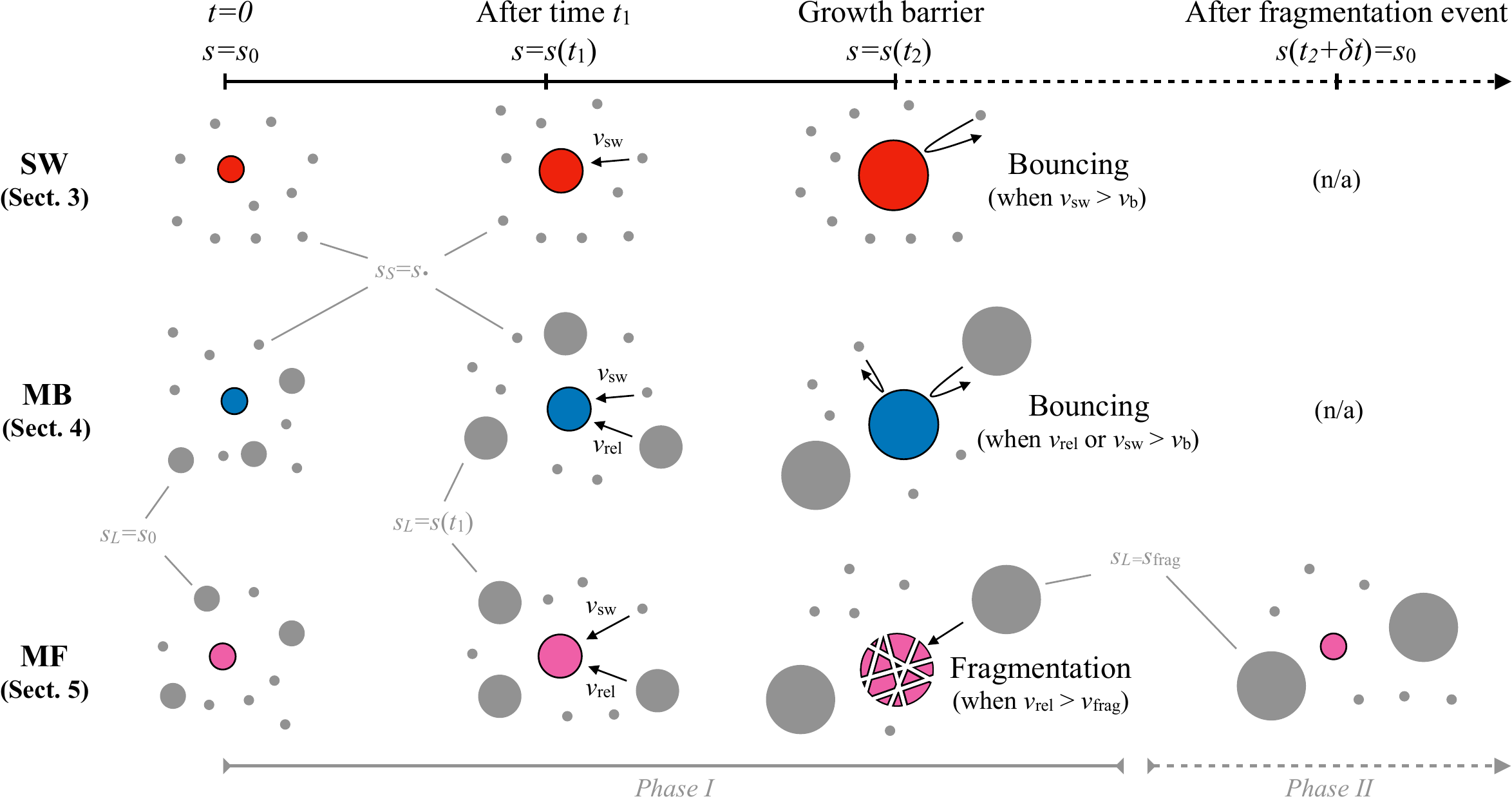}
\caption{Schematic illustrating the (local) dust coagulation process in the three different scenarios considered in this work. The particle that is being followed is shown in color and its size is given at the top of the Figure. Top row: In the sweep-up growth scenario (SW, Sect.~\ref{sec:smallgrowth}), the red particle grows by sweeping up small $s_\bullet$-sized background dust particles. Growth stalls due to bouncing when the sweep-up velocity $v_\mathrm{sw}>v_\mathrm{b}$. Middle row: In the mixed growth with bouncing case (MB, Sect.~\ref{sec:bouncing}), a fraction of the background dust population is always assumed to have the same size as the particle of interest, $s_L=s(t)$. Growth again stalls due to bouncing. Bottom row: The mixed growth with catastrophic fragmentation case (MF, Sect.~\ref{sec:fragmentation}) is similar (during Phase I) to the MB case in the sense that is has a co-evolving background dust population. In the MF scenario, however, bouncing does not occur and growth is instead halted by fragmentation (when $v_\mathrm{rel}>v_\mathrm{frag}$). After fragmentation has occurred (i.e., during Phase II), the particle we are following is reset to $s_0$ while the large bodies in the background population maintain a size $s_\mathrm{frag}$.}\label{fig:sketch}
\end{figure*}

Here we develop a framework for following individual grains as they travel radially and vertically in the disk while being incorporated into different-size aggregates via collisions. We consider a variety of growth scenarios, starting from a simplified model where dust exists as some non-evolving background population that is well-mixed throughout the gas and is swept up as the dust grain moves through the disk.  This serves to help to identify key evolutionary processes which control the dynamical evolution of the dust.  In order to capture the detailed behavior of dust coagulation in a disk, while keeping the problem more computationally tractable, we then extend this approach to consider the collisional evolution of dust grains with both small background dust and `like-sized' particles that co-evolve with the grain of interest.   These models are described in greater detail in the sections that follow and illustrated in Fig.~\ref{fig:sketch}.

\section{Transport with Sweep-Up}\label{sec:smallgrowth}

\subsection{Growth Model}
The first scenario we consider is similar to the simple one-particle sweep-up model that has been described by \citet{safronov1972} and \citet{dullemond2005} in the context of vertical settling. In this scenario, we simulate a single dust grain as it moves and accumulates mass by sweeping up small dust particles. The background dust population always and everywhere consists of $s_\bullet = 0.1~\mathrm{\mu m}$-size particles that are well-coupled to the gas (top row of Fig.~\ref{fig:sketch}). Thus, at every location the dust density is given by\footnote{We ignore the collisional trapping effect, which can skew the small grain abundance towards the midplane at disk radii where $f_\mathrm{d}$ is high \citep[e.g.,][Fig.~6]{krijtciesla2016}.} $\rho_\mathrm{d}/\rho_\mathrm{g}= \Sigma_\mathrm{d}/\Sigma_\mathrm{g}$, where $\Sigma_\mathrm{d}/\Sigma_\mathrm{g} \equiv f_\mathrm{d}=0.01$ equals the vertically integrated dust-to-gas ratio. The instantaneous growth rate of a particle with size $s \gg s_\bullet$ that is sweeping up background dust particles can then be written as
\begin{equation}\label{eq:dm_sw}
\left( \frac{\partial m}{\partial t} \right)_\mathrm{sw}= \pi s^{2} v_{\mathrm{sw}} \rho_\mathrm{d},
\end{equation}
where $v_\mathrm{sw}$ is the relative velocity between two particles of sizes $s$ and $s_\bullet$, which we calculate following \citet[][Eq.~16]{okuzumi2012} accounting for Brownian motion, differential settling, differential drift, and stochastic turbulence \citep{ormel2007b}.  Both $v_\mathrm{sw}$ and $\rho_\mathrm{d}$ are taken as the average values of the location of the particle at the beginning of the timestep (i.e., $r(t)$ and $z(t)$) and where it ends after the timestep (i.e., $r(t+\Delta t)$ and $z(t+\Delta t)$).  We assume all particles are spherical and remain at a constant material density $\rho_\bullet = 1.5\mathrm{~g~cm^{-3}}$, such that $m=(4/3)\pi s^3 \rho_\bullet$.

In a single simulation, we start with a grain with $s(t=0) = s_0 = 1\mathrm{~\mu m}$ (i.e., somewhat larger than the background dust population) initially at the midplane at $r=5$ AU from the star, the same initial conditions as described in Section \ref{sec:dynamics}. We then alternate between calculating the displacement of the particle over a timestep that is small compared to the local dynamical timescale (here we take  $\Delta t=(1/50)\times 2\pi /\Omega$) and mass gain according to Eq.~\ref{eq:dm_sw}.  The new location and size (or equivalently mass) of the grain's host particle are then recorded and serve as the starting conditions for the next timestep. A physical limit to growth is added in the form of a bouncing barrier: if $v_\mathrm{sw}$ exceeds a threshold velocity of $v_\mathrm{b}=0.5 \mathrm{~m/s}$, the growth rate is set to 0. The value for the bouncing threshold velocity is chosen based on the experiments of \citet{musiolik2016}, who found a bouncing threshold of $0.43\mathrm{~m/s}$ for ${\sim}100\mathrm{~\mu m}$-sized aggregates composed of a 1:1 mixture of $\mathrm{H_2O}$ and $\mathrm{CO_2}$ ice.  If the particle subsequently moves to a region where $v_\mathrm{sw} < v_\mathrm{b}$, further growth is again allowed according to Eq.~\ref{eq:dm_sw} until the barrier is reached again. 

The calculation is repeated until the grain reaches the inner edge of the disk, taken here to be 0.2 AU.  For each scenario, we perform $10^3$ simulations each tracking an individual grain from the same initial conditions in order to properly characterize the broad distribution of evolutionary histories that grains may experience within the protoplanetary disk.

\subsection{Grain Evolution}

The middle row of Figure~\ref{fig:individual} shows the evolution of three individual grains in our sweep-up simulations, plotting their paths through the disk, with the colors of the trajectories indicating their host aggregate's size. In their early evolution, the grains we follow are either free-floating or found within small aggregates, and thus their motions are dominated by turbulent diffusion.  As such, at early stages they exhibit motions similar to those shown in the top row of Fig.~\ref{fig:individual}, with each grain/aggregate following a unique path. 

Over time, however, the followed particle sweeps up more and more mass, leading to an increase in its size (and Stokes number).  As a result, vertical settling and radial drift begin to dominate (approximately when $\St > \alpha$, roughly at $s\sim0.01\mathrm{~cm}$).  This leads to  conditions that favor rapid growth as the particle settles near the midplane where dust densities are highest, providing more material to sweep up.  As they grow larger, the particles drift inwards more rapidly and are more and more confined to the midplane.  As a result, the fate for each particle is ultimately the same: they drift to the inner edge of the disk in ${\sim}10^5$ years.

\begin{figure}[t]
\centering
\includegraphics[width=.45\textwidth]{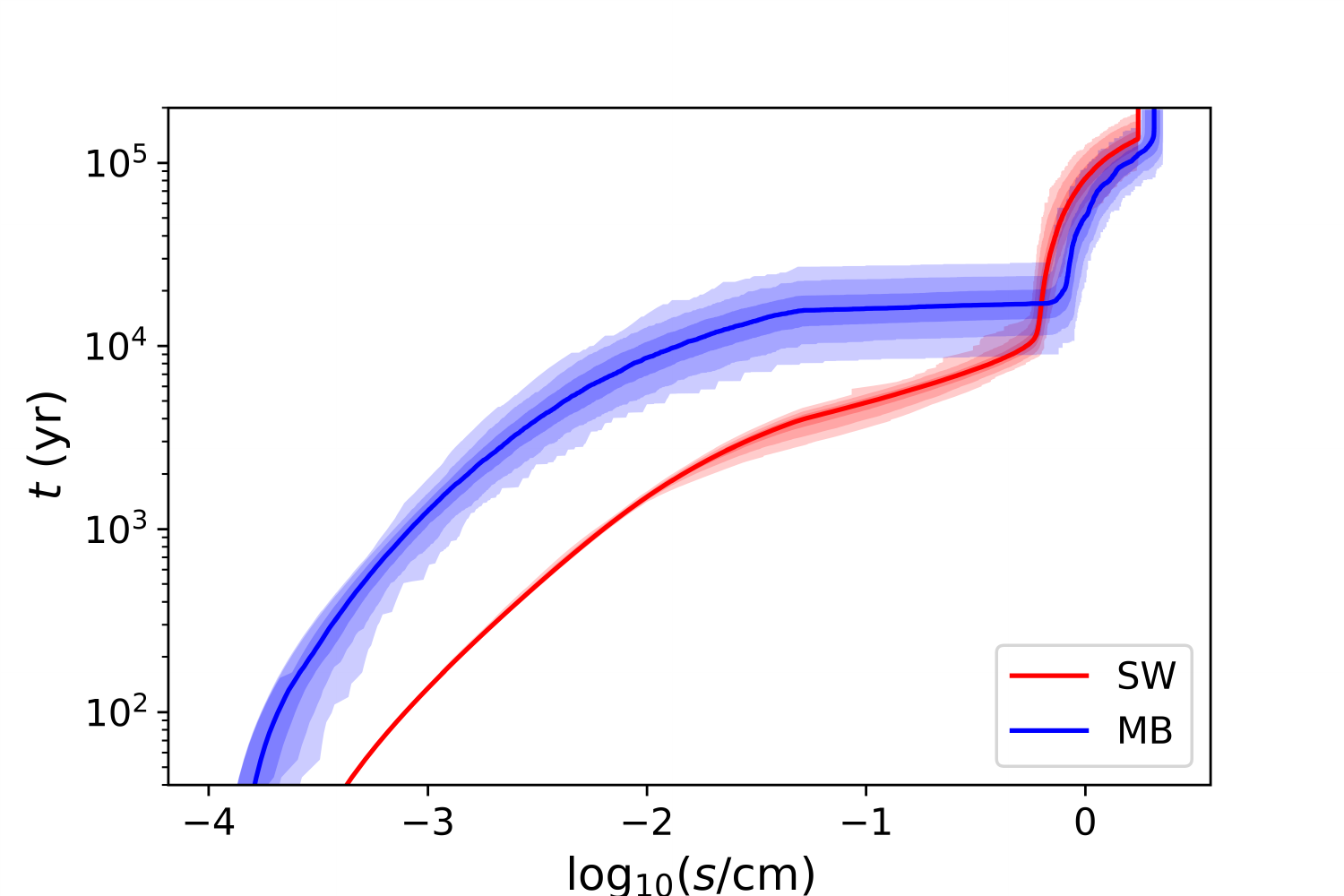}
\caption{Evolution of host aggregate size distribution with time of 1000 particles starting at $r=5$ AU in a disk with $\alpha=10^{-4}$. Solid lines denote the median size at each time, while shadings represent the range in which 68\%, 95\%, and 99.7\% of the aggregates fall. The sweep-up scenario is in red, and the bouncing-limited mixed growth scenario is in blue.}\label{fig:sizes}
\end{figure}

The red curve in Fig.~\ref{fig:sizes} summarizes the 1000 different simulations we performed for the conditions described here, showing the distribution of aggregate sizes containing the grains we followed over time.  The solid curve indicates the median value for different simulations, while the red shaded regions show the range of the distribution, marking where 68\%, 95\%, and 99.7\% of the aggregates fall\footnote{While the distribution is not necessarily Gaussian, we use these percentiles, which correspond to 1, 2, and 3 standard deviations, as guides to illustrate the variations in behavior.  Thus, the ranges show the minimum and maximum values of the "central" 680, 950, and 997 of the 1000 particles we track.}.  We see that this SW distribution is quite narrow, with 95\% of the grains being contained in aggregates that vary in size by no more than a factor of $\sim$2 at a given time.  A pronounced kink develops after $\sim$10$^{4}$ years, once the grains are in aggregates nearly 1 cm in size.  
This is a result of the bouncing which occurs when relative velocities are above 0.5 m/s: particles grow as long as the relative velocity is below this threshold, stopping once it reaches this critical value, and then continuing, temporarily, if the particle migrates into a region where the relative velocities drop below this threshold.  Growth thus slows dramatically for all particles once they reach this size. While each aggregate takes its own path through the disk, the final sizes of aggregates, as they arrive at the inner edge of the disk, cluster tightly around ${\sim}2\mathrm{~cm}$.  Importantly, such sizes are comparable to the dust particles observed in protoplanetary disks, as is the prediction that the maximum size of grains decreases with increasing distance from the star \citep[e.g.][]{tazzari2016}.

\begin{figure}[t]
\centering
\includegraphics[width=.45\textwidth]{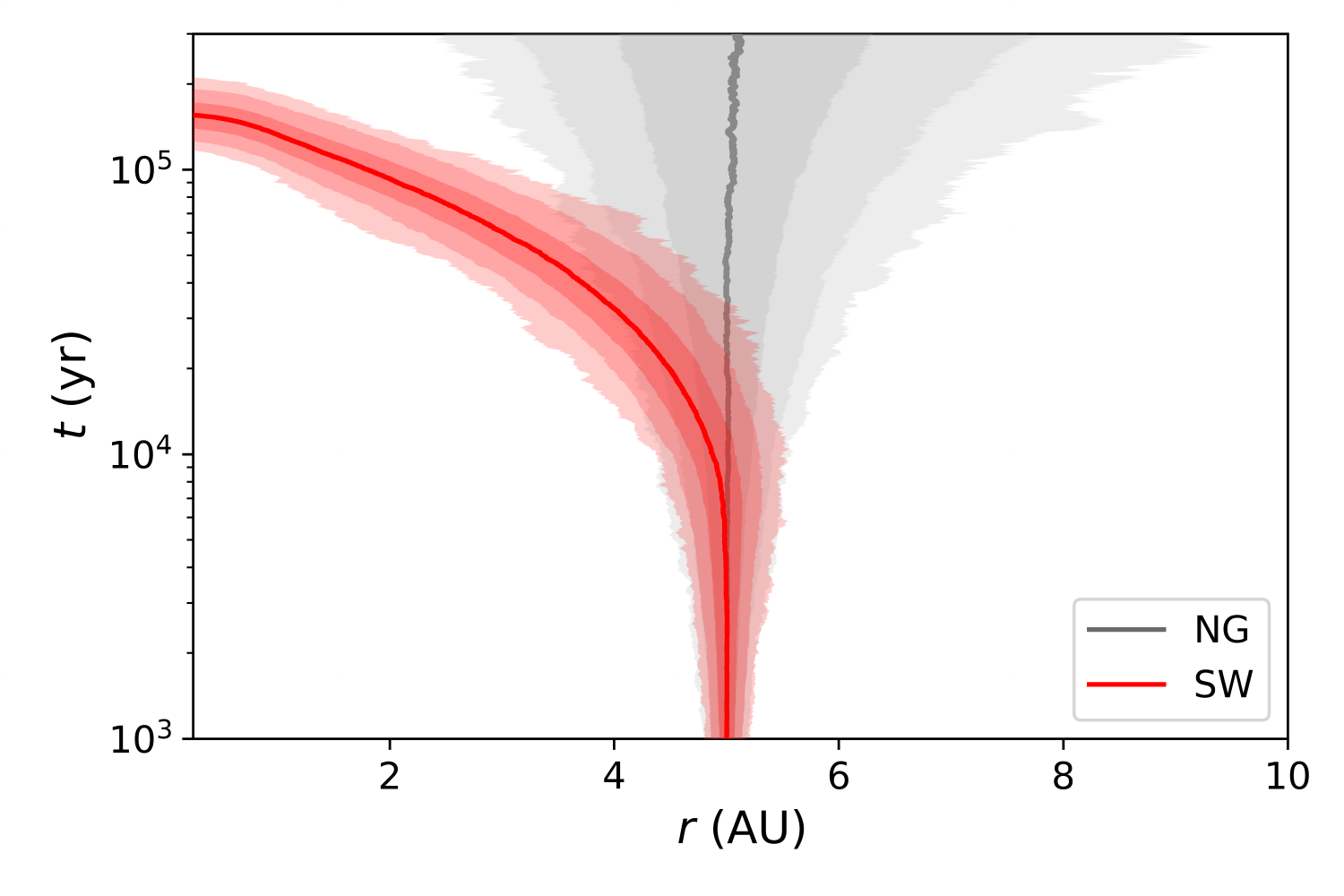}\\
\includegraphics[width=.45\textwidth]{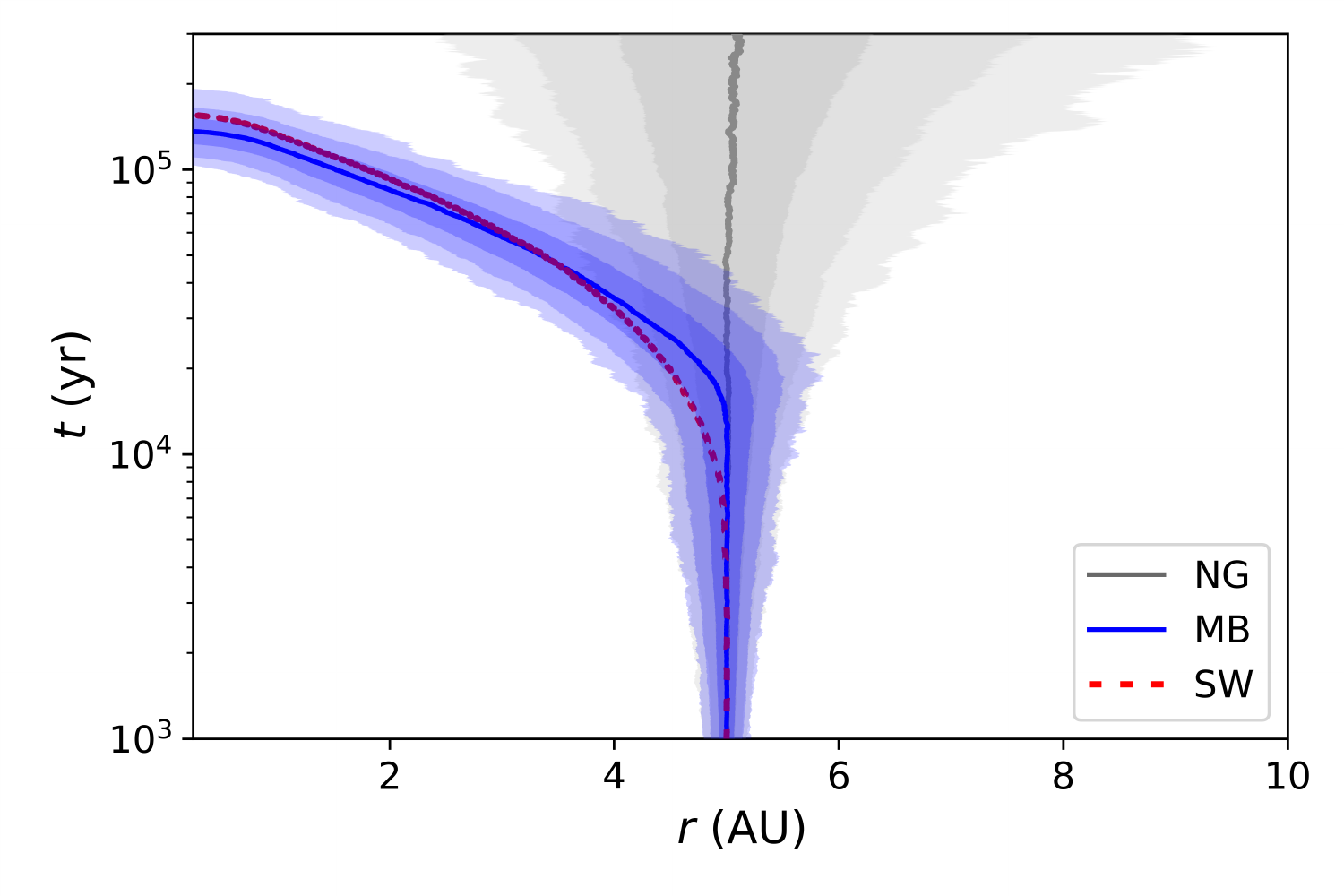}
\caption{Evolution of radial distribution of 1000 grains with time, with all particles starting at $r=5$ AU in a disk with $\alpha=10^{-4}$. Solid lines show the median location and shadings represent the range in which 68\%, 95\%, and 99.7\% of the aggregates fall, and the no growth radial distribution is plotted in gray in the background. Top: the sweep-up case (Sect.~\ref{sec:smallgrowth}). Bottom: the mixed growth with bouncing scenario (Sect.~\ref{sec:bouncing}), with the median line from the sweep-up case for comparison.}\label{fig:collective}
\end{figure}

While the final parent aggregate size and total time spent in the disk do not vary substantially among the 1000 grains considered here, we do see significant variations in the individual trajectories taken through the disk, particularly early in their evolution.  The top panel of Fig.~\ref{fig:collective} shows the distribution of radii at which the particles were found over time (the distribution without growth is shown in the background for reference), with the median behavior of the particles plotted as a line, along with shaded ranges showing the extent of how  68\% 95\%, and 99.7\% of the particles were spread at a given moment.  Again, during early periods, the grains are individual solids or part of small aggregates, with the dominant dynamic process operating on them being turbulent diffusion.  As such, the distribution of locations remain centered on the starting location of 5 AU, spreading outwards over time similarly to the no growth case.  This evolution lasts for a period of ${\lesssim}10^{4}$ years, after which the median distance from the star decreases steadily with time.  This change in behavior coincides roughly with when the particles reach sizes $s\sim0.01\mathrm{~cm}$, corresponding to $\mathrm{St} \sim 6 \times 10^{-4}$, or a few times $\alpha$, when they begin to decouple from the gas. From this point onward, vertical settling and later radial drift begin to dominate, with particles beginning their inevitable march to the inner edge of the disk.

\section{Transport with Mixed Growth: Bouncing}\label{sec:bouncing}
\subsection{Growth Model}
While the simple picture above shows how the dynamical evolution of a dust grain is controlled by the size of the aggregate it is contained in, dust growth is expected to be more complex than the simple sweep-up model considered above. In reality, the background dust population will also experience grain growth and settling, leading to changes in the size and spatial distributions of potential collision partners. 

Here, we develop a simplified model for a co-evolving distribution of collision partners to provide a more self-consistent model for how particle growth would occur. In developing our approach we follow \citet{birnstiel12} in separating the solids available for the particle of interest to interact with into two populations (middle row of Fig.~\ref{fig:sketch}): a \emph{Small} population which is well-coupled to the gas (with grain sizes $s_S=s_\bullet=0.1\mathrm{~\mu m}$) and a \emph{Large} population for which the size is always equal to that of the particle we are following (i.e., $s_L = s(t)$). At every disk radius $r$, a fraction $f_s$ of the local dust surface density is assumed to be present in the form of small grains. This leaves a fraction $(1-f_s)$ of the solids to evolve and grow at the exact same rate as the particle we are following. Here we set $f_{s}=0.1$, which falls between the values of $0.03$ and $0.25$ which are appropriate for drift-limited and fragmentation-limited local growth \citep{birnstiel12}. An aggregate of radius $s$ will then be able to collide with grains that are much smaller than it as well as grains that are comparable in size. Collisions with smaller particles will largely be continuous, allowing us to describe the growth from such collisions as done in the simple sweep-up model from the previous section (e.g., Eq.~\ref{eq:dm_sw}, with $\rho_\mathrm{d}$ replaced with $\rho_{S}$=$f_{s}  f_\mathrm{d} \rho_\mathrm{g}$). Growth via collisions with like-sized aggregates will occur through random, distinct events, resulting in the doubling of the aggregate's mass.
To accurately capture this behavior and the dynamics of the particles between these growth events, we expand our growth model to include a Monte Carlo approach similar to that used by \citet{krijtciesla2016}. 

In this \emph{mixed growth} approach, we first evaluate the typical collision time for impacts with similar-size bodies
\begin{equation}\label{eq:t_col}
t_\mathrm{col} = \frac{m}{4 \pi s^{2} v_\mathrm{rel} \rho_L},
\end{equation}
in which $v_\mathrm{rel}$ is the relative velocity between similar\footnote{To account for a finite width of the large dust size distribution, we evaluate $v_\mathrm{rel}$ as the relative velocity between particles of sizes $s$ and $(1+\varepsilon)s$, with $\varepsilon=0.1$.} particles, again following \citet[][Eq.~16]{okuzumi2012} as above and $\rho_L = (1-f_s) f_\mathrm{d} \Sigma / (\sqrt{2\pi}H_L) \exp( - 0.5 (z/H_L)^2 )$ is the local density of large aggregates. Vertical settling is included when calculating the scale height of the large particle population via $H_L = H \sqrt{\alpha / (\alpha+\St)}$ \citep[e.g.,][]{youdin2007}. We do not follow the radial motions of the background dust, maintaining a constant $f_\mathrm{d}=0.01$, a valid assumption if the disk is large and inward drifting solids are replaced by those further out over the short timescales considered here. Each timestep, we check if the particle experiences a large collision by first drawing a random number $\mathcal{R} \in [0,1]$ to determine the time $\Delta t_\mathrm{col}$ until the next large collision event
\begin{equation}\label{eq:dt_col}
\Delta t_\mathrm{col} = - t_\mathrm{col} \log \mathcal{R}.
\end{equation}
If $\Delta t_\mathrm{col} > \Delta t$, we can ignore similar-size impacts during this timestep and update the particle mass as
\begin{eqnarray}\label{eq:dm_bounce1}
m(t+\Delta t) =& m(t) + \left(\dfrac{\partial m}{\partial t}\right)_\mathrm{sw} \Delta t, \nonumber \\ 
=& m(t) + \pi s^{2} v_{\mathrm{sw}} \rho_S \Delta t.
\end{eqnarray}
Alternatively, if $\Delta t_\mathrm{col} \leq \Delta t$, we end the current timestep immediately after the mass-doubling impact (i.e., after $\Delta t_\mathrm{col}$) and set:
\begin{eqnarray}\label{eq:dm_bounce2}
m(t+\Delta t_\mathrm{col}) =& 2m(t) + \left(\dfrac{\partial m}{\partial t}\right)_\mathrm{sw} \Delta t_\mathrm{col}, \nonumber \\
=& 2m(t) + \pi s^{2} v_{\mathrm{sw}} \rho_S \Delta t_\mathrm{col}.
\end{eqnarray}
This use of Eq.~\ref{eq:dt_col} in combination with a random number to find the time until the next collision event, follows the Monte Carlo methodology described by \citet{ormel2007}, who, in turn, based their approach on that derived by \citet{gillespie75}. This method introduces a natural stochasticity to the local collisional growth process while capturing the expected growth rate over many timesteps. The dynamics are handled exactly as in Sect.~\ref{sec:smallgrowth}, and we re-evaluate Eqs. \ref{eq:t_col} and \ref{eq:dt_col}, including drawing a new random number $\mathcal{R}$, at the beginning of every timestep to take into account the particle's new location and to robustly sample the probability density function.

Finally, as in the previous Section, we modify the growth equations to include the effect of bouncing, setting $(\partial m / \partial t)_\mathrm{sw}=0$ and/or subtracting $m(t)$ from the r.h.s. of Eq.~\ref{eq:dm_bounce2} when $v_\mathrm{sw}>v_\mathrm{b}$ and/or $v_\mathrm{rel} > v_\mathrm{b}$, respectively. We note here that, because relative velocities are functions of both particle sizes, the bouncing barriers for sweep-up growth and mass-doubling growth may be reached at (slightly) different points.

With collisions (be it with small or like-sized bodies) resulting in either sticking or bouncing, the size of the dust grain we are following through the disk is a monotonically increasing function of time: $(\partial s / \partial t) \geq 0$. Collisional fragmentation (for which this is no longer the case) will replace the bouncing barrier in the cases considered in Sect.~\ref{sec:fragmentation}.

\subsection{Grain Evolution}

Again, we simulated the evolution of 1000 different particles as they were subjected to the transport and growth as described above.  The bottom row of Fig.~\ref{fig:individual} shows examples of trajectories followed by the dust grains in our model, while Figs.~\ref{fig:sizes} and \ref{fig:collective} show the distributions in size and radial location of the mixed growth particles with comparison to the no-growth and sweep-up scenarios.

Differences in growth timescales are readily seen among the growth scenarios in Fig.~\ref{fig:sizes}. At small particle sizes, growth in our mixed case takes longer than in the pure sweep-up case.  This is because collision velocities, and therefore collision rates, are low between well-coupled similar-sized particles \citep[e.g.,][]{okuzumi2012}. Thus, growth in a given timestep early on typically occurs through sweep-up of the small population, which is present only at 10\% of the abundance as in the pure sweep-up model ($f_{s}$=0.1 here). Once aggregates grow to larger sizes ($\gtrsim$0.01 cm), two effects combine to increase the rate of growth of these aggregates. First, $v_\mathrm{rel}$ increases significantly as particles decouple from the smallest turbulent eddies \citep{ormel2007b}, leading to more frequent mass-doubling collisions. Second, the local density $\rho_L$ in the midplane increases as large solids settle and become more concentrated.

The stochastic nature of the similar-size collisions increases the spread in carrier aggregate sizes at any given time, especially during the period of rapid growth outlined above.  This is shown in the span of particle sizes present at a given time in Figure 3, with the greatest range around $t \sim 10^{4}$ yr, when the grains we follow are in aggregates ranging from 0.1 mm to nearly 1 cm.  Once particles reach the upper portion of this range, they attain relative velocities that are near the bouncing barrier.  As before, growth then generally proceeds much more slowly, only as the particle drifts into regions where the relative velocity the particle attains allows for solids to stick instead of bounce. 

Ultimately, the fate of the grains in this MB scenario is the same as that for sweep up: the grains are contained within aggregates that drift inside of 0.2 AU on a timescale of $\sim$10$^{5}$ years.  This is illustrated by comparing the radial distribution of solids over time for the two growth scenarios shown in Figure \ref{fig:collective}.  Overall the two curves show similar behavior, with the mixed growth grains undergoing a sharper transition from diffusion-dominated evolution (following the no-growth background) to drift-dominated, a result of faster growth at these sizes due to like-sized collisions.

\begin{figure*}[t]
\centering
\includegraphics[width=.5\textwidth]{colormap_rev.pdf}
\\
\includegraphics[width=.33\textwidth]{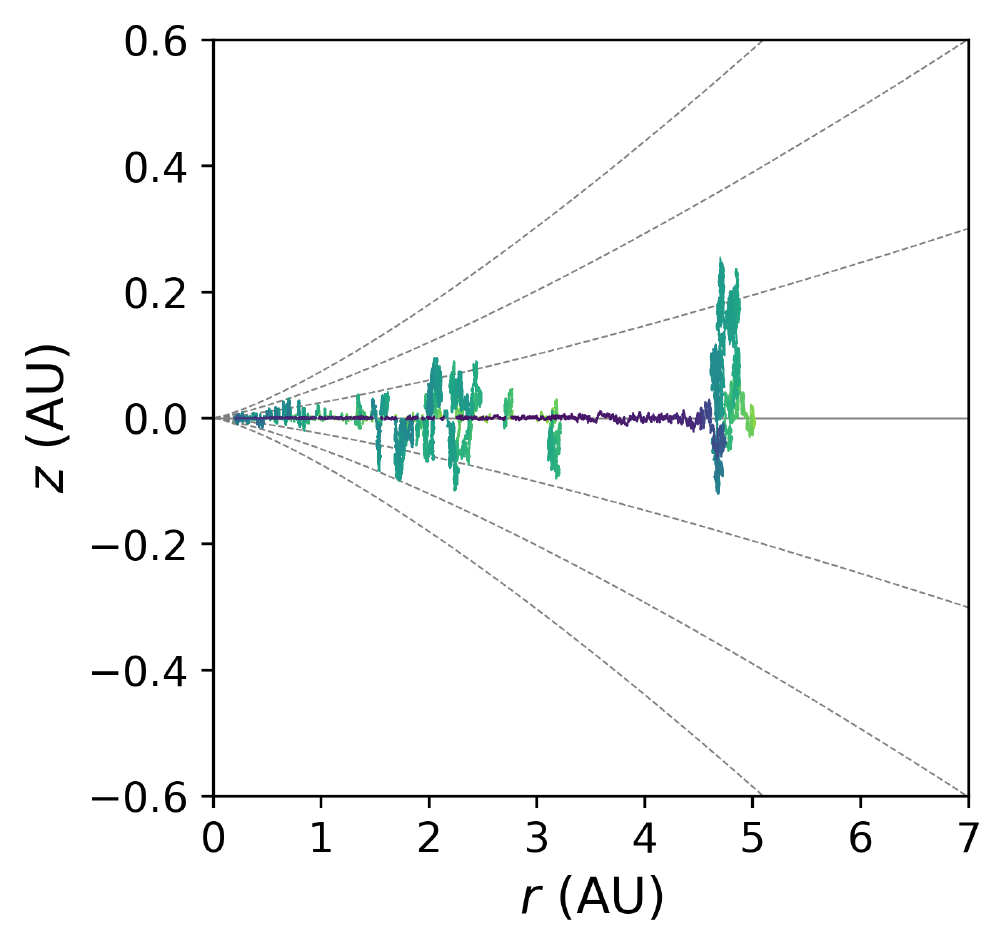}
\includegraphics[width=.33\textwidth]{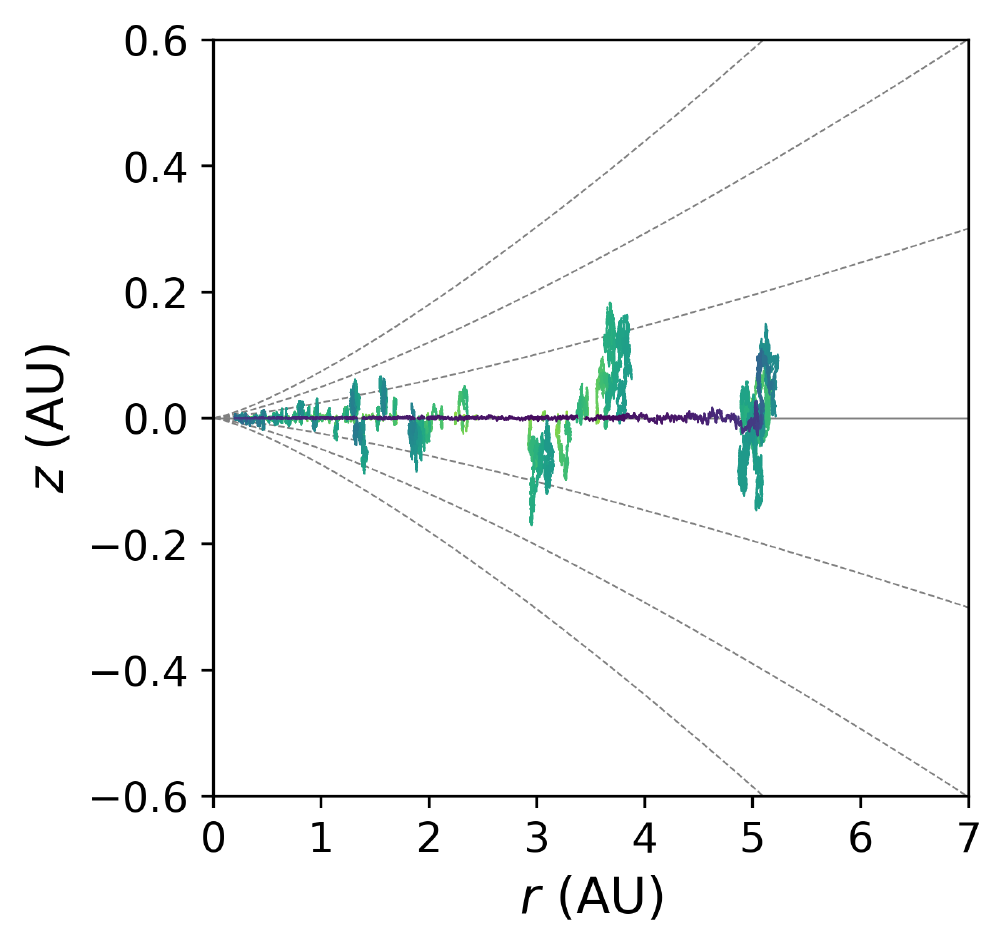}
\includegraphics[width=.33\textwidth]{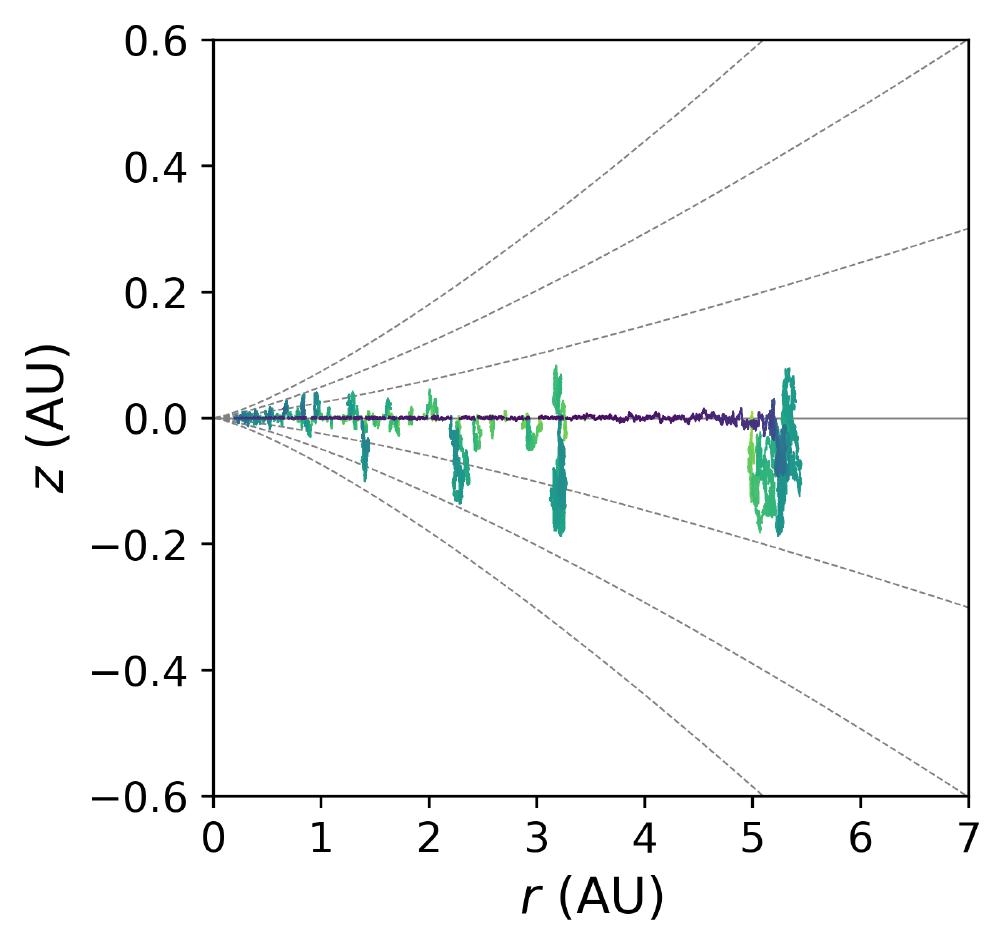}
\caption{Same as Fig.~\ref{fig:individual}, but for the scenario with catastrophic fragmentation included (Sect.~\ref{sec:fragmentation}). }\label{fig:individual2}
\end{figure*}

\begin{figure}[t]
\centering
\includegraphics[width=.45\textwidth]{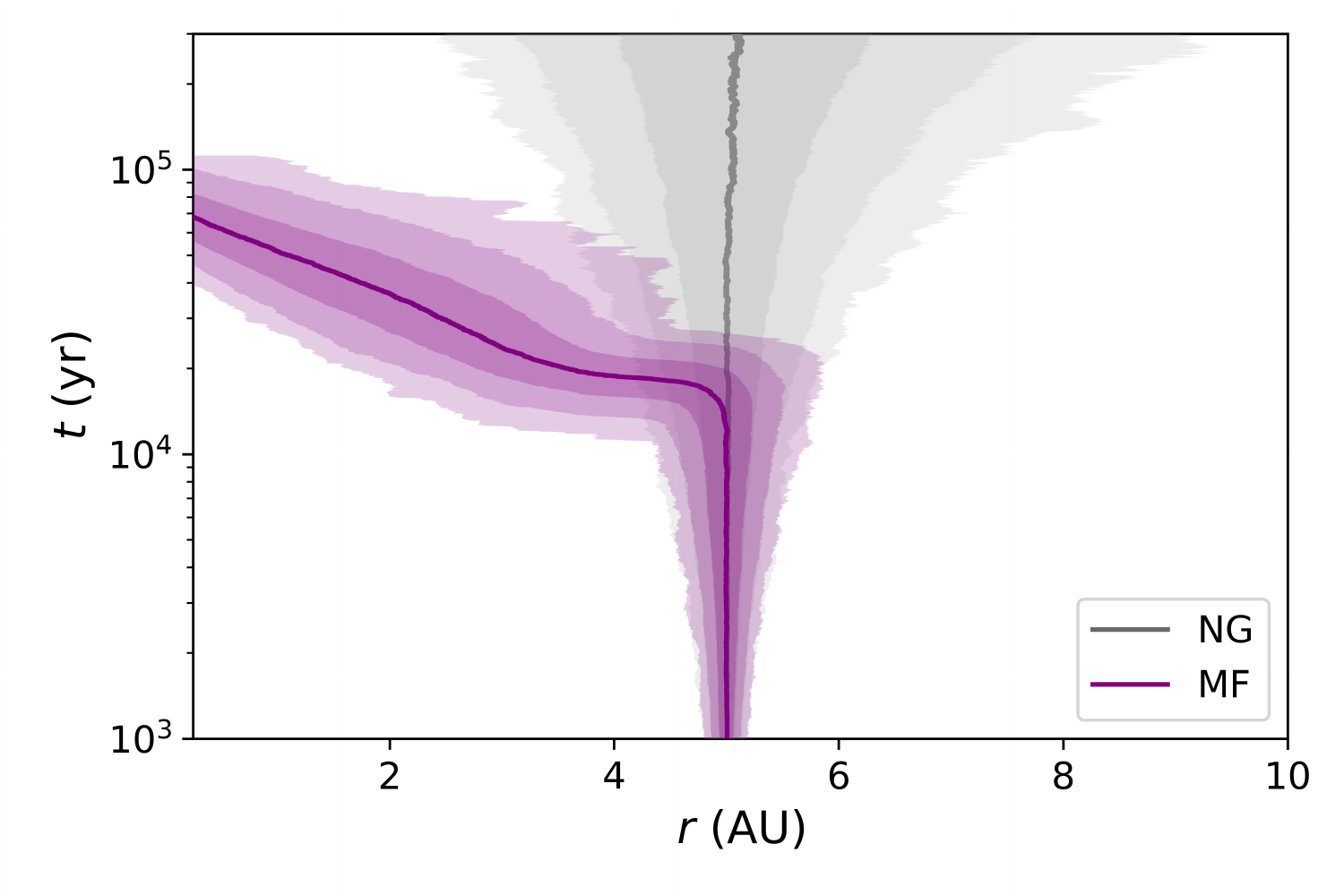}
\caption{Similar to Fig.~\ref{fig:collective}, but for growth with fragmentation (Sect.~\ref{sec:fragmentation}).}\label{fig:collective_frag}
\end{figure}

\section{Transport with Mixed Growth: Fragmentation}\label{sec:fragmentation}

\subsection{Growth model}\label{sec:fragmentation_setup}
In many studies of coagulation, fragmentation is considered to be the limiting factor of growth rather than bouncing \citep[e.g.,][]{brauer2008,birnstiel2010}.  In this picture, as solids grow and encounter one another at greater velocities, the energies cause the collision partners to break apart rather than merge or survive.  As a result, when aggregates collide at velocity $v_\mathrm{rel}>v_\mathrm{frag}$, their mass is redistributed as fragments that replenish the small grain population \citep{dullemond2005}. In local models of dust coagulation, the outcome is usually a growth/fragmentation steady-state distribution of particle sizes, with a small fraction of the total mass contained in small dust, and a large fraction concentrated around a maximum size above which collisions lead to aggregate destruction \citep[e.g.,][]{birnstiel2011}.

The value of $v_\mathrm{frag}$ varies with aggregate composition and structure, with typical values ranging from ${\sim}1\mathrm{~m/s}$ for refractory materials to ${\gtrsim}10\mathrm{~m/s}$ for water ice grains \citep{dominiktielens1997,wada2013,gundlach2015}, although these high fragmentation velocities have recently been called into question as the presence of other ices or very low temperatures may reduce the stickiness of ice-covered grains \citep{musiolik2016,musiolik2019}. We adopt a constant and relatively conservative value of $v_\mathrm{frag}=5\mathrm{~m/s}$ and discuss the effects that different values would have on our results below in Sect. \ref{sec:parameterexp}.

In defining the behavior of dust grains in this mixed growth with fragmentation (MF) scenario, we identify two phases of growth (bottom row of Fig.~\ref{fig:sketch}):
\emph{Phase I} is similar to growth in the previous section, replacing the bouncing barrier $v_\mathrm{b}$ with a fragmentation barrier $v_\mathrm{frag}$, and represents an early phase where dust grains and the particles that they interact with are small and have yet to establish the steady-state distribution.  During this phase, collisions among particles are such that $v_\mathrm{rel}<v_\mathrm{frag}$, allowing larger aggregates to form.  As such, we can make use of Eq.~\ref{eq:dm_bounce1} and Eq.~\ref{eq:dm_bounce2} to describe grain growth. This phase lasts until the aggregate containing the grain we are following experiences a collision with another aggregate at a velocity $v_\mathrm{rel} > v_\mathrm{frag}$, in which case fragmentation occurs. We assume that in this event the aggregate is completely disrupted, and the grain we are following returns to being a free-floating particle with size $s_0=1\mathrm{~\mu m}$.  

Once a particle is released in a fragmentation collision, we transition to
\emph{Phase II}, where the particle we are following can still collide with a distribution of background solids. However, the sizes of solids are assumed to be set by coagulation-fragmentation equilibrium and maintain a new, bimodal steady-state distribution of particle sizes.  There is still a population of sub-micron sized particles that the grain can sweep-up, as in Phase I and as considered in our other growth scenarios.  However, the largest grains are no longer the same size as the particle we are following ($s(t)$): instead their size corresponds to the local fragmentation-limited size\footnote{This size is found by equating $v_\mathrm{rel}=v_\mathrm{frag}$ for two like-sized particles, assuming turbulence is the dominant velocity source \citep{birnstiel2011}.} $s_\mathrm{frag}$ (see Fig.~\ref{fig:sketch}). The procedure is then similar to that described for the MB case in the Section above: if no large collision occurs, the particle only sweeps up small grains, and Eq.~\ref{eq:dm_bounce1} is used. If a collision with a larger particle takes place, we calculate the new particle mass as:
\begin{eqnarray}\label{eq:dm_frag}
m(t+\Delta t_\mathrm{col}) = 
\begin{cases}
m(t) + m_\mathrm{frag} +\left(\dfrac{\partial m}{\partial t}\right)_\mathrm{sw} \Delta t_\mathrm{col} &\textrm{~if~} v_\mathrm{rel} < v_\mathrm{frag},\\
m_0 &\textrm{~if~} v_\mathrm{rel} \geq v_\mathrm{frag},
\end{cases}
\end{eqnarray}
where $m_\mathrm{frag}$ and $m_0$ are masses corresponding to $s_\mathrm{frag}$ and $s_0$, respectively. The first case in Eq.~\ref{eq:dm_frag} corresponds to the case where the particle we are following collides and sticks to an aggregate of mass $m_\mathrm{frag}$ (the local fragmentation limit). The second case in Eq.~\ref{eq:dm_frag} corresponds to a fragmentation event, after which the particle returns to having a size $s_0$. Since very small grains cannot reasonably cause the catastrophic fragmentation of large aggregates, we add an additional constraint to Eq.~\ref{eq:dm_frag}, only allowing fragmentation to occur when the mass ratio of the colliders ${>}10^{-2}$, assuming small particles to be accreted/swept-up otherwise\footnote{We ignore here the process of erosion \citep{seizinger2013b,schrapler2018}.}.

We note that having the particle return to its original size after each fragmentation event is an extreme scenario; in reality, fragments of a range of sizes are likely to develop, from the monomer we consider up to a bit smaller than $s_{\mathrm{frag}}$ \citep[e.g.,][]{birnstiel2011}.  If grains were instead contained in these larger fragments, they would continue to drift inwards and grow more rapidly than we consider here.  Therefore, the lifetimes of grains in the disk considered here likely represent upper limits, with most actual grains drifting inwards more rapidly and being lost on shorter timescales.

Thus, the overall evolutionary scenario we follow here is that a grain begins with size $s_0$ at $t=0$ and initially grows by sweeping up small grains and sticking to like-sized aggregates (Phase I) until it experiences a destructive collision and is released as a small fragment. From that point onward, the grain grows via sweep-up or by sticking to a large aggregate of size $s_\mathrm{frag}$ (Phase II). During Phase II, many growth/fragmentation cycles can take place before the grain of interest reaches the inner disk.

\subsection{Grain Evolution}
Examples of individual grain trajectories in the fragmentation case are shown in  Fig.~\ref{fig:individual2}. As before, we set the fraction of small grains such that $f_{s}=0.1$. Grains migrate above the disk midplane early, as expected for small grains lofted by turbulence, but these excursions to high altitudes are limited, as growth occurs and particles are incorporated into larger aggregates that settle to the disk midplane and then drift inwards.  This is similar to the previous growth cases we considered here, though the settling to the midplane is more extreme as grains can grow to larger sizes here (as $v_{\mathrm{frag}} > v_{\mathrm{b}}$). That is, in a typical fragmentation case considered here, particles reach sizes of $\sim$50 cm near 5 AU, which corresponds to a Stokes number $\mathrm{St} \approx 0.3$ and a large particle scale height $H_L/H \approx 0.02$.  For comparison, in the bouncing scenario (Sect.~\ref{sec:bouncing}), particles reach sizes of only $s=1$ cm near 5 AU, corresponding to $\mathrm{St} \approx 6 \times 10^{-3}$ and $H_L/H \approx 0.13$.

A key difference in the overall trajectories seen here, however, is that the grains experience multiple distinct excursions to higher altitudes before eventually reaching the inner disk.  This is a result of the grains being liberated by fragmenting collisions involving the large aggregates that they are contained within.  These collisions release the grains as small aggregates once again, allowing them to diffuse away from the midplane.  Eventually the grains are re-incorporated into larger aggregates, and the process repeats.  A similar cycling of grains in and out of larger aggregates was discussed in \citet{krijtciesla2016}, who also found that collisional growth increases the amount of time a particle spends near the disk midplane and reduces its total residence time at higher altitudes.

Fig.~\ref{fig:collective_frag}  plots the evolution of the radial location of a swarm of 1000 grains released at the midplane at 5 AU with the same fixed fragmentation velocity $v_\mathrm{frag}=5\mathrm{~m/s}$.  A similar evolution is seen as in the previous growth cases, with the early redistribution of solids occurring as in the no growth cases, as some particles move closer to the star and others diffuse further away.  Eventually, though, growth leads
to radial drift dominating, with particles migrating towards the star, where they are ultimately lost. This transition occurs a bit more dramatically here than in our previous runs, as the collection of particles shifts inwards rapidly after $\sim$10$^{4}$ years, with grains moving almost horizontally on the plot at this time. This is again because the drifting aggregates reach larger sizes here than in our previous cases, allowing grains to grow to higher Stokes numbers than in our bouncing regime (0.3 vs 6$\times 10^{-3}$) and resulting in drift rates of ${\sim}3\times10^{-3}\mathrm{~AU/yr}$ vs $7\times10^{-5}\mathrm{~AU/yr}$, respectively. As such, the grains drift inwards much faster, with the average particle reaching the edge of the disk on shorter timescales ($<$10$^{5}$ years) relative to the previous scenarios. 

\begin{figure}[t]
\centering
\includegraphics[width=.45\textwidth]{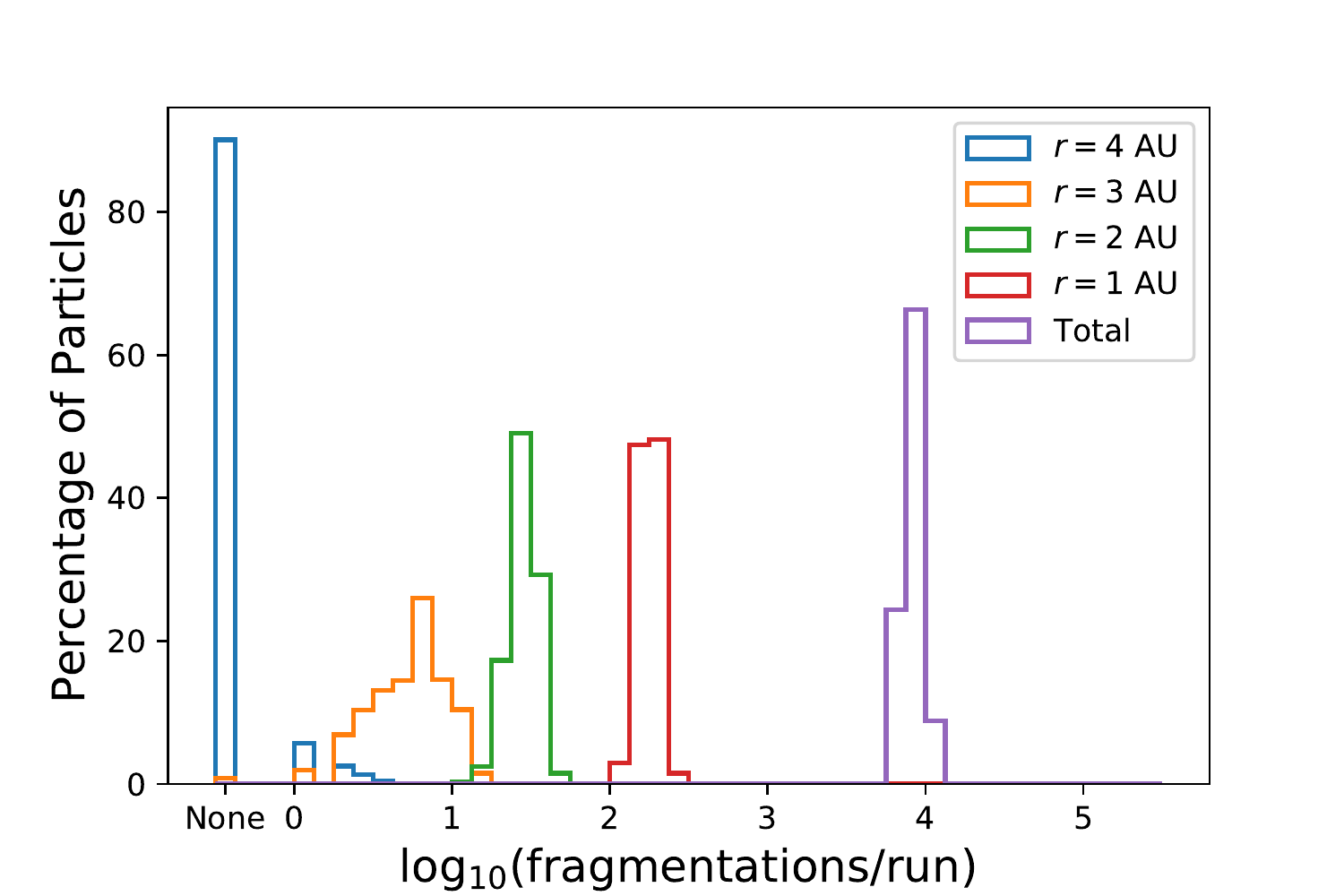}
\caption{Distribution of number of fragmentations experienced by particles when first crossing the specified radius, with total number of fragmentations per run also shown. Particles start from $r=5$ AU with fragmentation threshold velocity $v_{\mathrm{frag}}=5$ m/s.}\label{fig:frag_counts}
\end{figure}

 Fig.~\ref{fig:frag_counts} shows the distribution of the total number of fragmentation (liberation) events that grains experienced at the time when they \emph{first} crossed specific radial locations. Nearly 90\% of these grains reach $r=4\mathrm{~AU}$ before being liberated in a fragmentation event and entering Phase II of growth, while a small fraction ($\sim$10\%) had been involved in multiple fragmentation events before reaching 4 AU for the first time. As grains continue to migrate inwards, the typical number of fragmenting collisions that they were involved in increases by orders of magnitude.  This is both because of the higher densities in this region and the higher relative velocities.  This effect is seen in Fig. 4, as the number of excursions away from the midplane (when the grains are liberated as small particles after a collision) increases at smaller heliocentric distance.  This is also consistent with simulations that focus on grain distributions (rather than individual trajectories), which find that fragmentation dominates in the inner few AU, while particle growth is drift-limited at larger radii \citep[e.g.,][]{birnstiel12}.

\begin{figure*}[t]
\centering
\includegraphics[width=.45\textwidth]{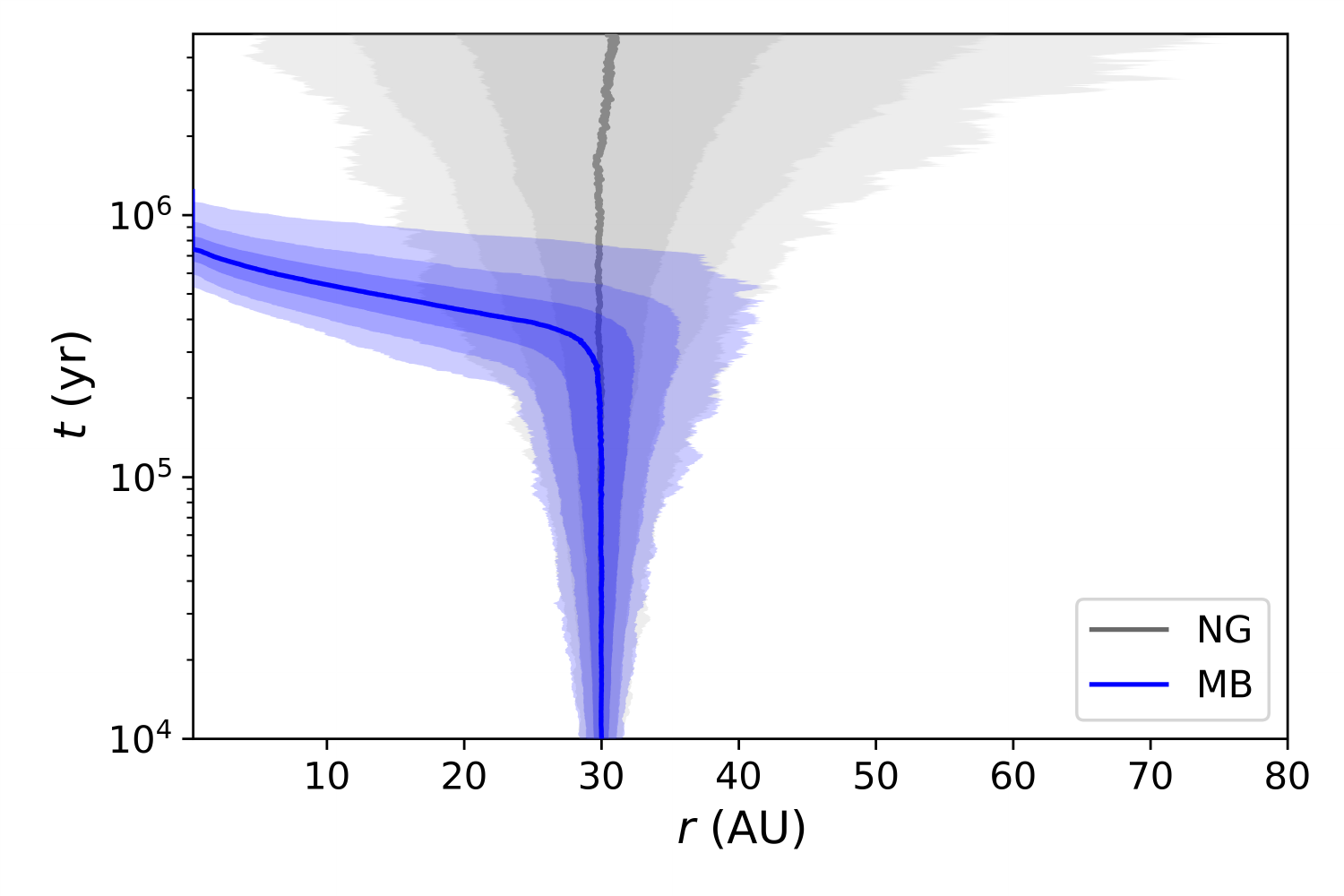}~~~
\includegraphics[width=.45\textwidth]{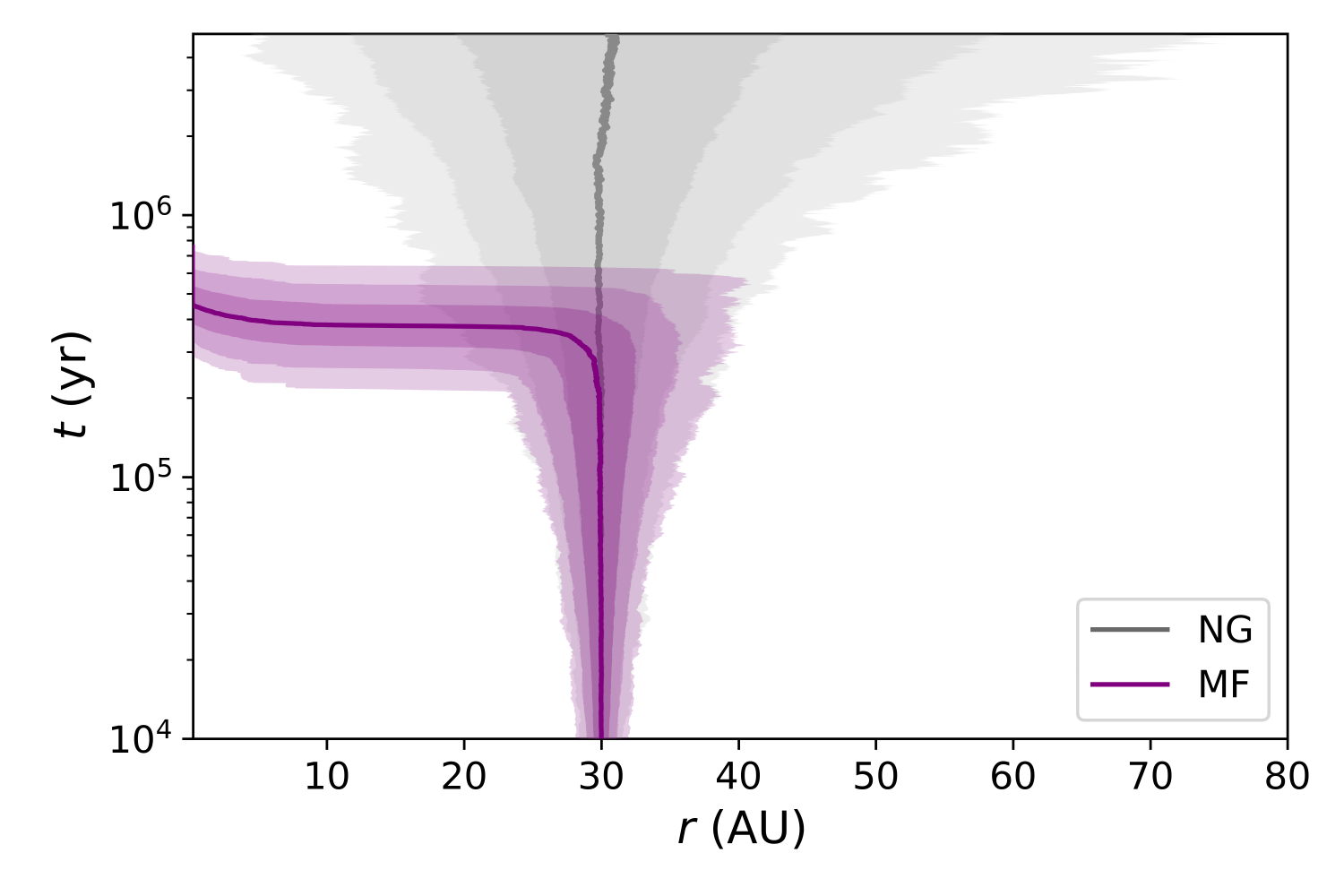}

\caption{Similar to Figs.~\ref{fig:collective} and \ref{fig:collective_frag}, but for dust grains starting out at $r=30\mathrm{~AU}$. Left: Mixed growth with bouncing at $v_\mathrm{bounce}=0.5\mathrm{~m/s}$. Right: Mixed growth with fragmentation at $v_\mathrm{frag}=5\mathrm{~m/s}$.}\label{fig:collective_radii_2}
\end{figure*}

\begin{deluxetable*}{ l | c c c c | c c c c }[t]
\centering
\tablecaption{Summary of model results.}
\tablewidth{0pt}
\tablehead{
\colhead{} & \multicolumn{4}{c}{Standard} &  \multicolumn{4}{c}{$r_0=30\mathrm{~AU}$}\\
\colhead{} & \colhead{NG} & \colhead{SW} & \colhead{MB} & \colhead{MF} & \colhead{NG} & \colhead{SW} & \colhead{MB} & \colhead{MF}
}
\startdata
Sect. & \ref{sec:dynamics} & \ref{sec:smallgrowth} & \ref{sec:bouncing} & \ref{sec:fragmentation} & \ref{sec:disc} & \ref{sec:disc} & \ref{sec:disc} & \ref{sec:disc} \\
$t_{50\%}/\mathrm{kyr}$ & ${>}t_\mathrm{end}$ & 155.7 & 136.6 & 69.1 & ${>}t_\mathrm{end}$ & 709 & 740 & 453 \\
$t_{2\%}/\mathrm{kyr}$ & ${>}t_\mathrm{end}$ & 125.9 & 110.7 & 47.3 & ${>}t_\mathrm{end}$ & 601 & 592 & 329 \\
$ s_{\mathrm{final}} / \mathrm{cm}$ & $10^{-4}$ & 1.72-1.74 & 1.79-2.27 & $10^{-4}$-4.91 & $10^{-4}$ & 1.71-1.74 & 1.78-2.27 & $10^{-4}$-4.89 \\
$\langle s_{\mathrm{final}} \rangle_{\mathrm{m}} / \mathrm{cm}$ & $10^{-4}$ & 1.74 & 2.07 & 4.78 & $10^{-4}$ & 1.74 & 2.06 & 4.78 \\
$R_{2\%}^\mathrm{max}/ \mathrm{AU}$ & 7.87 & 5.33 & 5.53 & 5.53 & 59.7 & 33.5 & 35.9 & 36.1 \\
 $t_{R_{2\%}^{\mathrm{max}}}/ \mathrm{kyr}$ & $t_\mathrm{end}$ & 8.3 & 18.6 & 16.9 & $t_\mathrm{end}$ & 123 & 390 & 328\\
\enddata
\tablecomments{The standard case corresponds to $\alpha=10^{-4}$ with grains starting at $r_0=5\mathrm{~AU}$. NG = No growth; SW = Sweep-up growth; MB = Mixed growth with bouncing; MF = Mixed growth with fragmentation.}
\label{tab:summary}
\end{deluxetable*}

\section{Dependence on Model Parameters}\label{sec:parameterexp}

In all cases considered here, the coupled growth and dynamical evolution of the grains we followed in the protoplanetary disk were the same: they grew to sizes that led to settling and inward drift, resulting in the grains released at 5 AU being lost at the inner edge of the disk on time scales that are short compared to the lifetime of the protoplanetary disk.  This common outcome was a result of growth occurring such that grains were efficiently incorporated into, and then spent much of their lifetimes within, aggregates whose dynamics were dominated by gas drag-induced drift.

The specific outcomes shown here, and by extension the histories of the grains within the disk, are particularly sensitive to the physics of how grains behave in collisions and what limits their growth. In particular, the results depend sensitively on the choice of the limiting velocity at which grains bounce or fragment ($v_{\mathrm{b}}$ and $v_{\mathrm{frag}}$). We have employed constant, universal values of $v_\mathrm{b}=0.5\mathrm{~m/s}$ (Sects.~\ref{sec:smallgrowth} and \ref{sec:bouncing}) and $v_\mathrm{frag}=5\mathrm{~m/s}$ (Sect.~\ref{sec:fragmentation}). Theoretical and experimental studies indicate these threshold velocities depend on particle composition, and therefore, disk location. In particular, the presence of water ice mantles is believed to increase the stickiness of grains (Sect.~\ref{sec:fragmentation_setup}), resulting in sharp drops in the sticking efficiency (and maximum sizes) of grains inside the water snowline \citep[e.g.,][]{banzatti2015} where water ice is not stable. In the context of our simulations, the absence of water ice in the region where $T\gtrsim150\mathrm{K}$ would result in overall smaller particles and a reduced drift rate inside $r\lesssim1\mathrm{~AU}$. Specifically, the radial drift velocity scales as $v_\mathrm{drift} \propto \St \propto v_\mathrm{f}^{2}$ (or ${\propto} v_\mathrm{b}^{2}$ for bouncing) when turbulence dominates the relative velocities \citep{birnstiel12}, suggesting that the timescale on which radial drift operates can be increased by 1-2 orders of magnitude inside the snowline, resulting in a local pile-up of small grains \citep{banzatti2015}. Conversely, the dust-to-gas ratio and particle size can become elevated just outside the water snowline by the condensation of water vapor coming from the inner disk \citep{ros2013,drazkowska2017,schoonenberg2017}. While these effects were not included in our calculations, we would still expect limited outward mixing in the region exterior to the snowline as discussed above. Further, the level to which water ice-coated grains could survive collisions has recently been questioned as \citet{musiolik2016} and \citet{musiolik2019} suggest that the presence of other ice species and the low temperatures expected in disks could reduce their stickiness.  As a result, $v_{\mathrm{frag}}$ may be lower in the outer regions of protoplanetary disks, resulting in smaller maximum sizes of icy aggregates. This would likely lead to longer survival times of the dust grains in the disk \citep[e.g.,][]{birnstiel2010}, though our lower values of $v_\mathrm{b}$ show that ultimately the fates would be similar.

The level of turbulence (set by $\alpha$ in our disk model) in the disk also plays an important role. This parameter defines the level of stirring that dust grains will experience from the gas, and thus it is a key factor in setting the relative velocities with which grains move with respect to one another (once grains grow beyond a few tens of microns).  Lower levels of turbulence will yield more gentle collisions, and thus larger aggregates and fewer fragmentation events. On the other hand, the lower collision speeds can increase the duration of the initial growth phase \citep[e.g.,][]{sato2016}. Once large aggregates eventually form, the degree of vertical settling and the velocity with which they drift will be higher in disks with a reduced turbulence strength.

The density structure of the gaseous disk will also impact the evolution of the dust grains. As particle dynamics are controlled by their interactions with the gas, more massive disks than considered here would result in grains having to grow to larger physical sizes to begin settling and drifting (i.e., to reach the same Stokes number).  However, the higher mass would also increase the spatial density of dust, and thus the rate of growth of particles.  These effects essentially offset one another, leading to similar outcomes as seen here.  Lower dust-to-gas ratios, meanwhile, would result in slower growth, increasing the lifetimes of particles within the disk \citep[e.g.,][Eq.~13]{birnstiel12}.

The process of radial drift is in turn sensitive to the pressure gradient in the disk midplane, with large grains drifting slower in regions where the pressure gradient is smaller \citep{weidenschilling1977}. In the inner disk, the existence of a viscously heated region can alter the pressure gradient \citep[e.g.,][Eq.~29]{ida2016}, but the effect will be relatively small and limited to the region $r\lesssim1\mathrm{~AU}$ unless the disk accretion rate exceeds $10^{-8}~M_\odot/\mathrm{yr}$. Further out in the disk, the presence of pressure maxima in the outer disk can halt drift altogether and keep particles from migrating inward \citep{pinilla12}.  We return to this point in Sect.~\ref{sec:disc}.

Lastly, if dust aggregates are more porous or fractal than considered here, then their growth to large Stokes numbers will be slower than what has been discussed thus far.  Such aggregates would remain coupled to the gas more effectively, following the dynamical evolution seen when treating solids as individual, small grains \citep{ormel2007,okuzumi2012,krijtormeletal2016}. However aggregates must grow to the point such that planetesimals are able to form, something that generally requires particles with $ \St \gtrsim 10^{-3}$ to be efficient \citep{cuzzi01,cuzzi08,youdin05,carrera15}.  The extent to which grains form and are able to remain as fractal aggregates is an area of ongoing research.

\section{Implications for Chemical Mixing in the Disk}\label{sec:disc}

Table \ref{tab:summary} provides a suite of statistical measures of the
behavior of the dust grain histories for the various growth conditions
described here.  These include $R_{2\%}^\mathrm{max}$, the largest distance from the star within which all but 2.2\% of particles were located at any time in the simulation (i.e., reaching beyond this distance was a rare outcome). Grains in the no-growth case reach distances beyond 7 AU in the ${\sim}10^{5}$ years studied here, but in all growth cases, outward movement is severely limited, with the most extreme grains typically migrating out to only $5.3-5.5$ AU before drifting inwards again and eventually being lost to the star.  What did vary across the growth regimes, however, was the time at which the most outward diffusing particles reached their largest distances from the star, ($t_{R_{2\%}^{\mathrm{max}}}$). In the sweep-up case, the extreme diffusers began their march inwards after $\sim$8000 years, while both mixed growth cases took over 2$\times$ longer, allowing grains to diffuse out a bit further as evidenced by the larger values of $R_{2\%}^\mathrm{max}$. This is due to the fact that growth at small sizes was faster in the sweep-up case (as illustrated in Figure \ref{fig:sizes}), leading to the grains getting incorporated into larger particles that begin to settle and drift inwards earlier than in the mixed growth cases.  However, continued growth from sweep-up is relatively slow compared to the mixed growth cases, as collisions with like-sized particles lead to much more rapid growth above $\sim$0.01 cm.  Thus particles in the two mixed growth cases grow much more rapidly beyond this size, and drift inwards toward the star at much more rapid rates.

Grains are most rapidly lost by drifting to the inner edge of the disk in the fragmentation case, with the timescale for 50\% of the grains being lost ($t_{50\%}$) being just over 69 kyr, with bouncing-limited mixed growth and sweep-up timescales being about 2$\times$ and 2.25$\times$ longer.  We also quantified the dispersion of the particle behavior, finding when the first 2.2\% of grains are lost ($t_{2\%}$), with results being similarly ordered, meaning that the distributions of behaviors of grains did not differ significantly across the growth cases. Also provided in Table \ref{tab:summary} are the range and mass-weighted average of the aggregates' size when they reach the inner edge of the disk ($s_\mathrm{final}$ and $\langle s_{\mathrm{final}} \rangle_{\mathrm{m}}$ respectively). As discussed above, these final sizes are constrained by the different growth barriers, and the stochastic nature of the final collisions leads to the spread in possible final sizes in the bouncing- and fragmentation-limited mixed growth cases, though in the latter case most of the mass reaching the inner disk is in the larger particles.
Thus, despite the different growth scenarios and uncertainties in model parameters, a robust finding in all our simulations is that dust growth and dynamics are intimately coupled and combine to control how materials are redistributed in a protoplanetary disk.  In all growth cases, these two processes lead to rapid inward movement of solids on timescales that are short compared to disk lifetimes.

Such a finding is important when looking at the properties of primitive materials in our own Solar System, such as the outward transport of chondritic-like fine-grained (micron-sized) materials to the comet formation region \citep{brownlee06,mckeegan06} or the preservation of millimeter to centimeter-sized CAIs within the disk for millions of years \citep{krot09,connelly12}.
 While turbulent diffusion \citep[e.g.][]{bockelee02,cuzzi03,ciesla2010,jacquet11,desch2018} or stirring in gravitationally unstable disks \citep[e.g.][]{boss08} 
could deliver solids that began in the inner solar nebula to the region where comets would form, in the absence of other effects, particle growth would have significantly offset this transport, limiting the efficacy by which such materials were delivered.

This discussion would also hold if delivery of high-temperature grains from the inner disk to the outer disk occurred through disk winds, where MHD effects lead to gas being carried away from the disk along magnetic field lines \citep[e.g.][]{konigl10,bai2016}.  It has been suggested that solids could be entrained in these winds, falling back onto the disk at larger heliocentric distances than where they began \citep{safier93,giacalone19}.  We  investigated this scenario by running similar models as those outlined above, but for grains starting at 30 AU instead of 5 AU.  The key statistics for these runs are also given in Table \ref{tab:summary}, with the time evolution of the radial locations of the grains within our bouncing and fragmentation limited growth cases shown in Figure \ref{fig:collective_radii_2}.  A similar behavior is readily seen as before, with grains diffusing around within small aggregates early, before transitioning to a drift-dominated regime leading to their migration and loss to the inner edge of the disk in $<$10$^{6}$ yrs.  Thus solids delivered to larger heliocentric distances via winds would still migrate back to the inner disk again on relatively short timescales.

This rapid inward movement of materials means that exchange and communication across large radial distances could be quite effective in protoplanetary disks.  However, \citet{kruijer2017} suggested that distinct Carbonaceous Chondrite (CC) and Non-Carbonaceous (NC) reservoirs existed in the solar nebula, resulting in planetesimals forming with characteristic isotopic abundances over an extended period of time, and thus that communication or transport between these reservoirs was limited.  They interpreted this dichotomy as arising from the growth of Jupiter within the solar nebula after 1 Myr and it preventing materials from being exchanged across its orbit, thus preserving the distinct reservoirs. Our results, however, suggest that materials can be exchanged over a radial span of 5-20 AU on  timescales of only ${\sim}10^{5}$ years.  If different isotopic reservoirs exist at different radial locations, then a barrier, such as a pressure bump or gap in the disk, must develop on this timescale to prevent communication between them.

One way to offset the rapid loss of grains via radial drift as we have found here would be to limit growth.  As discussed above, if the threshold for reaching the bouncing or fragmentation barrier were lower (lower $v_{\mathrm{b}}$ or $v_{\mathrm{frag}}$), then solids would remain within small aggregates, reducing the effect of drift.  This would also be achieved in more turbulent disks, as the velocity barriers would be reached by smaller particles, again limiting the role that drift could play.  However, more turbulence and longer lifetimes as smaller particles would enable more efficient mixing of solids via diffusion, which could be problematic in preserving distinct isotopic reservoirs as discussed above.  

 The process that stalls dust growth, whether bouncing or fragmentation, will also impact how well we detect signatures of mixing.    As shown in comparing Figures \ref{fig:sizes} and \ref{fig:collective}, growth to the bouncing barrier occurs in a timescale of $\sim$10$^{4}$ years, over which time the grains have not diffused far from their original location.  As such, the drifting aggregates will largely be composed of dust grains that originated in close proximity to one another.  In the case of fragmentation, however, as shown in Figure \ref{fig:frag_counts}, grains see multiple collisions which allow them to be released as smaller components and then incorporated into new aggregates again.  This cycling in and out of larger aggregates throughout the period of radial drift will lead to grains becoming mixed with materials that originated over a large range of locations in the disk.  This would be particularly true in the inner disk where fragmenting collisions are more common; in the drift dominated regime of the outer disk, those aggregates may be made of a more homogeneous collection of materials.

This discussion thus far, however, has ignored other dynamical effects that may be present in disks.  Radial expansion of the disk due to viscous evolution can enhance the outward transport of materials in the disk \citep{cuzzi03,jacquet11}.  Furthermore, viscous disks can develop flow structures where the movement of the gas is a function of height; in some cases, the midplane may see outward flows develop, pushing solids in this region outward or at least offsetting the inward migration rates that develop from gas drag \citep[e.g.][]{takeuchilin02,ciesla07,desch2018}.  These effects would have to be very strong and provide outward gaseous flows whose velocities are comparable to the inward drift rate of the solids that form here to be effective at preventing the inward migration of drifting solids. \citet{takeuchilin02} found that outward flows along the midplane of $\sim$10$^{-4}$ AU/yr, nearly enough to completely offset even our largest drifting particles considered here, would develop in a purely viscous disk for the disk structure assumed here.  Whether such conditions develop or are long-lived enough to offset the inward drift of solids needs to be investigated.
  
Finally, the modeling presented here assumed a smooth disk structure, with the pressure gradient decreasing monotonically with distance.  It is this pressure gradient that drives radial drift and the inward migration of solids over time. Recent observations have shown that rings and spiral structures are present in protoplanetary disks, even at very early stages of evolution \citep{andrews18}. Such structures have been hypothesized to be pressure bumps or gaps in the disk, where gas pressures reach local maxima or minima, allowing dust to concentrate in these regions and preventing their further inward migration \citep{pinilla12,dullemond18}.  While such structures may be due to presence of planets within the disk \citep{zhang18}, alternative means of creating these bumps have also been proposed that rely on hydrodynamic effects \citep[e.g.,][]{flock15}. These structures would serve as important barriers to the inward drift of materials, possibly working to prevent the loss of materials as seen here, but again would have to develop early in order to prevent the rapid loss of solids from the disk, or at least large-scale inward mixing of outer nebular materials.

\section{Summary}\label{sec:summary}

Here we have investigated the coupled transport and growth of solids in a protoplanetary disk.  We have studied various scenarios for dust growth, including bouncing and fragmentation barriers, and found that growth via coagulation has an important effect on how solids migrate and are redistributed within a given disk.  Our major findings are as follows:

\begin{itemize}

\item Dust growth and transport are intimately coupled.  While individual dust grains are readily redistributed by turbulence within a disk, collisions with other grains will also occur.  As these early collisions will lead to growth, the dynamics of the grains are set by the aggregates within which they are contained, leading them to settle to the disk midplane and drift towards the star (e.g., Figs.~\ref{fig:individual} and \ref{fig:individual2}).  This limits the extent to which materials are able to diffuse outwards from their original location, with particles making it at most to distances  $\sim$10\% beyond their original location when growth occurs, as opposed to 60-70\% when growth is ignored (Figs.~\ref{fig:collective}, \ref{fig:collective_frag}, and Table~\ref{tab:summary}).  This is a robust result, regardless of whether growth is limited by fragmentation or bouncing.

\item Dust growth that is controlled by a bouncing barrier will largely lead to aggregates that form from a collection of dust that originated at similar locations.  When growth is hindered by a fragmentation barrier, this will lead to many cycles of grains being incorporated into larger aggregates and then liberated again as free particles (Fig.~\ref{fig:frag_counts}), and aggregates will therefore represent a greater mix of dust from different locations.

\item This inward drift would lead to rapid delivery of outer disk materials to the inner disk, allowing for rapid and efficient exchange of solids over large distances, with solids ultimately being lost to the inner edge of the disk.  Particles in our fiducial models were typically lost from 5 AU in less than 10$^{5}$ years, while particles from 30 AU were lost in less than 10$^{6}$ years (cf. Figs.~\ref{fig:collective}, \ref{fig:collective_frag}, and \ref{fig:collective_radii_2}). In our Solar System, this presents a challenge for preserving CAIs and distinct isotopic reservoirs separated in space for longer than 10$^{6}$ years as suggested by meteorite studies.  That such features are observed in our Solar System may indicate that outward flows, pressure bumps, or gaps were present in the solar nebula, preventing the ultimate fates found here.

\end{itemize}

\acknowledgments
We are very grateful to an anonymous referee who provided insightful and stimulating comments on an earlier draft of this manuscript.  The paper is greatly improved as a result. FC acknowledges support from NASA awards NNX14AQ17G and NNX15AD94G. SK acknowledges support through Hubble Fellowship Program HST-HF2-51394.002-A, provided by NASA through a grant from the Space Telescope Science Institute, which is operated by the Association of Universities for Research in Astronomy, Inc., under NASA contract NAS 5-26555. This material is also based upon work supported by the National Aeronautics and Space Administration under Agreement NNX15AD94G for the program ``Earths in Other Solar Systems.'' The results reported herein benefited from collaborations and/or information exchange within NASA's Nexus for Exoplanet System Science (NExSS) research coordination network sponsored by NASA's Science Mission Directorate.

\software{Numpy \citep{numpy}, Matplotlib \citep{hunter07}}

\bibliographystyle{yahapj}
\bibliography{dust_refs}

\end{document}